\documentclass[aps,epsfig,twocolumn,showpacs,email,amssymb,nobibnotes]{revtex4}
\usepackage{graphicx,color,amsmath,amssymb,psfrag,epsfig,multirow,slashbox}

\usepackage{chemarr}
\usepackage{empheq}
\providecommand{\e}[1]{\ensuremath{\cdot 10^{#1}}}
\usepackage{footnote}

\begin{document}
\title{Stochastic single-gene auto-regulation}

\author{Tom\'as Aquino$^1$, Elsa Abranches$^2$ and Ana Nunes$^1$ \\ 
}

\affiliation{$^1$Centro de F{\'\i}sica da Mat{\'e}ria Condensada and 
Departamento de F{\'\i}sica, Faculdade de Ci{\^e}ncias da Universidade de 
Lisboa, P-1649-003 Lisboa Codex, Portugal \\
}

\affiliation{$^2$Instituto de Medicina Molecular and Instituto de Histologia e Biologia do Desenvolvimento, Faculdade de Medicina da Universidade de Lisboa, Av. Prof. Egas Moniz, 1649-028 Lisboa, Portugal and Champalimaud Neuroscience Programme, at Instituto Gulbenkian de Ci\^{e}ncia, Rua da Quinta Grande, 6, 2780-156 Oeiras, Portugal}

\begin{abstract}
A detailed stochastic model of single-gene auto-regulation is established 
and  its solutions are explored when mRNA dynamics is
fast compared with protein dynamics and in the opposite regime. 
The model includes all the sources of randomness that are intrinsic to
the auto-regulation process and it considers both transcriptional and post-transcriptional
regulation. The timescale separation
allows the derivation of analytic expressions for the equilibrium 
distributions of protein and mRNA. These distributions are generally well described in the continuous approximation, which is used to discuss the qualitative features of the protein equilibrium distributions as a function of the biological parameters in the fast mRNA regime.
The performance of the timescale approximation is 
assessed by comparison with simulations of the full stochastic system, and a good quantitative
agreement is found for a wide range of parameter values. We show that either unimodal or
bimodal equilibrium protein distributions can arise, and we discuss the auto-regulation
mechanisms associated with bimodality.
\end{abstract}

\maketitle


\section{Introduction}
\label{intro}

The role of stochasticity in cells and microorganisms has been discussed theoretically since the 1970s~\cite{berg1977physics,berg1978model}.
Because cellular processes often rely on chemical reactions, and correspondingly on chance encounters 
between molecules or molecular complexes, stochastic effects due to small numbers are ubiquitous in the 
cell. In particular, cellular decision processes, which are of of paramount importance as 
they allow cells to react to the internal and external media, are based on gene activation and 
regulation, often depending on random association and dissociation events.
While many works focus on the limits imposed by stochasticity and the evolution of noise-minimization 
strategies~\cite{berg1977physics,bialek2005physical,bialek2008cooperativity,fraser2004noise}, there 
is a growing interest in possible functional roles of noise.
Generically, the basic role of randomness in gene expression is to provide a natural means of generating 
phenotype variability across a population, enhancing its capacity to quickly adapt to fast-changing conditions.

 The evolution of experimental molecular biology 
techniques has made single-cell measurements possible and brought numerous confirmations of the 
presence of stochastic effects in gene expression~\cite{elowitz2002stochastic}, prompting a renewed 
interest in the 
mechanisms underlying gene expression and regulation in general, and specifically on the sources of 
randomness affecting them. The fact that genes coding for specific proteins are often present in single 
copies may introduce considerable noise.  Furthermore,  mRNAs are commonly present in low copy numbers, 
from a few to a few hundred 
molecules, and many proteins also exist in low number. Because transcription, translation and 
degradation events are stochastic, finite size fluctuations in mRNA and protein numbers become 
important.  Stochastic effects 
may suffice to drive long excursions of a gene's expression to higher or lower values, producing well-defined pulses in single cell protein abundances over time and/or multimodal protein expression 
distributions in a population. 
Fluctuations of the biological parameters of the system under consideration are another source  of 
randomness. For example, we  
characterize an active gene by a constant effective transcription rate, while this rate  
may depend on the presence of transcription factors whose concentration fluctuations induce 
fluctuations of the effective rate. Examples of theoretical 
approaches to these ideas can be found in~\cite{friedman2006linking,kalmar2009regulated,rue2011gene}.

The recent  development of 
single-molecule techniques led to the experimental identification of another, more specific source 
of variability in gene expression that accounts for the heavy-tailed distributions
often found in measures of population distributions of protein and mRNA abundance: both transcription 
and translation have been found, in many cases, to occur in time-localized bursts resulting in a 
geometrically distributed number of molecules, 
see~\cite{cai2006stochastic,kaufmann2007stochastic,ozbudak2002regulation,taniguchi2010quantifying}.

As experimental evidence of these sources of randomness accumulates 
\cite{suter2011mammalian,larson2011realtime}, the tools of statistical physics are being called upon for 
the development of a theoretical understanding of 
the underlying mechanics in noisy gene expression. Several models of the simplest 
elements of a gene regulatory network have been studied as stochastic processes that include a 
representation of some of these sources  of randomness 
\cite{hornos2005selfregulating,friedman2006linking,mugler2009spectral,iyer-biswas2009stochasticity, 
assaf2011determining}. As expected, the 
stationary solutions of these models may differ significantly from what one would obtain by simply 
adding a noise term to the equations stemming from a deterministic description. Moreover, the analytic 
solutions that can be obtained under certain assumptions were found to be  
in agreement with a wide set of experimental data \cite{taniguchi2010quantifying}.

In this paper, we make use of these tools to 
study a bottom-up model for single-gene auto-regulation that includes all the sources of randomness that 
are intrinsic to the auto-regulation process and is applicable in general to any 
protein species, auto-regulated by means either of transcriptional, as it is commonly considered,
or post-transcriptional regulation. Analytic solutions of the general model obtained 
in two complementary approximations for the relative timescales of protein and mRNA dynamics
are discussed in terms of the qualitative features of the equilibrium protein distributions.
The conditions for these approximations to hold are studied in some detail, and 
in their expected region of validity we find
good quantitative agreement with the results of stochastic simulations of the full system. We use the analytic solutions
to discuss the conditions under which single gene auto-regulation gives rise to 
bimodal protein distributions. Although these distributions are often associated in the literature
with the presence of more complex regulation mechanisms, we find that those conditions
are quite general.

The paper is organized as follows. In Section~\ref{Model}, we establish the stochastic model. In Section~\ref{Solanal} we present the solutions of the model for the protein and mRNA equilibrium 
distributions in the timescale separation approximations, and we discuss the qualitative features of 
the former. In Section~\ref{Approx}, we study the validity of the
approximations and compare the approximate analytic solutions with the results of simulations. We 
conclude in Section~\ref{conclude}.
Six appendices contain technical details which are too cumbersome to include in the main text.


\section{Model}
\label{Model}

\begin{figure}[h]
\includegraphics[scale=0.7]{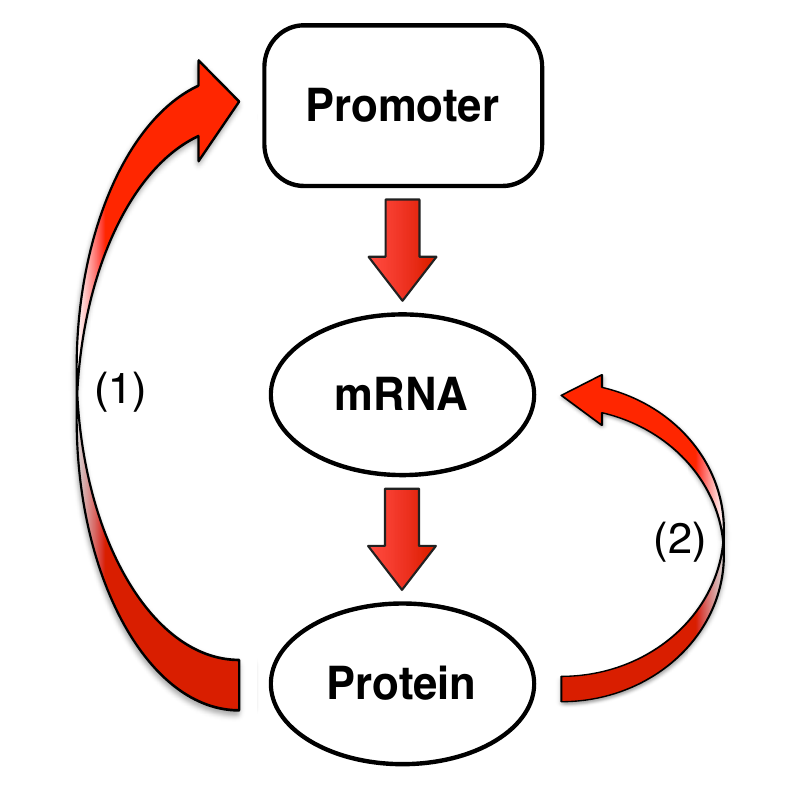}
\caption[Auto-regulation schemes]{(Color online) Basic structure of the dynamics of a 
single protein that auto-regulates either (1) transcriptionally or (2) translationally.
Arrows representing protein and mRNA degradation have been omitted.
}
\label{Fi:scheme}
\end{figure}

We study the cell-level dynamics, and corresponding population distributions, of a 
single protein capable of auto-regulation and its mRNA. 
Protein and mRNA concentrations
are controlled by the balance between production and degradation events. 
In transcriptional regulation (see Figure~\ref{Fi:scheme}, left arrow), the regulatory feed-back is mediated by binding of a molecule, whose concentration depends on that
of the protein itself, to the promoter region in the DNA to alter the transcription rate of its mRNA. This is the   
most commonly studied mechanism of gene regulation, but other mechanisms have been 
reported in the recent literature that act post-transcription, at the mRNA rather than at the promoter 
level \cite{waters2009regulatory}. In this translational regulation scenario, the regulator molecule interacts with the mRNA to change its rate of protein production (see 
Figure~\ref{Fi:scheme}, right arrow). In what follows we derive the Master Equations that govern protein 
and mRNA 
abundances in both these 
scenarios, starting with transcriptional regulation.

For concreteness, we will consider regulation to be effected by protein dimers, in agreement with experimental evidence for some particular
proteins \cite{wang2008requirement}.
Other choices, such as monomer binding \cite{hornos2005selfregulating} or a general 
cooperative binding modeled by a Hill 
function \cite{friedman2006linking} have been used in the literature. 
We assume the protein and mRNA populations to be 
non-interacting except for the fact that proteins dimerize prior to binding to the promoter.
The timescale of promoter reactions ($\sim$ seconds) is assumed much shorter than that of the mRNA 
($\sim$ minutes to hours) and protein (usually $\sim$ hours), in agreement with 
data for the typical timescales of these processes, 
see for example~\cite{bialek2005physical,schwanhausser2011global}.
Since regulation takes place at the DNA level, we need to model processes at three different levels: the promoter's (DNA), the mRNA's,  and the protein's. 

\subsection{DNA Level}

For promoter dynamics, we essentially follow~\cite{guantes2008multistable}, adapted to a fully 
stochastic description. The promoter site is assumed to bind only one dimer molecule at a time. 
To avoid unnecessarily heavy notation, in what follows we assume a given value of protein copy number 
$n$ in the cell; probabilities should accordingly be taken as conditional probabilities given $n$.
Denote by $P_{ f }$ the free promoter state and by $P_{ b }$ the bound state of the promoter and a 
dimer. For each instant $t$, let $p( P_{ f } , t )$ be the probability of the promoter being free, and 
$p( P_{ b } , t )$ the probability of it being bound to a dimer. 
 The evolution of the probability of the 
bound state is governed by the Master Equation
\begin{equation}
\label{PMeq}
	\dot{ p }( P_{ b } , t \mid j ) =
		j \ k^{ + }  
			p( P_{ f } , t \mid j ) -
		k^{ - } p( P_{ b } , t \mid j ) \ ,
\end{equation}
where $k^{ + }$ and $k^{ - }$ are the promoter site binding and unbinding rates, and 
$j$ is the number of dimer molecules available for binding to the promoter.
Since at all times the promoter is either free or bound to a dimer, we have also
$p( P_{ f } , t \mid j ) +	p( P_{ b } , t \mid j  ) = 1$.

The number of dimers in the cell as a function of protein copy number  
is given in a rate equation description by (see for instance \cite{van1992stochastic} or
Appendix~\ref{app:Dimers}):
\begin{equation}
	n_2(n) = \frac{n}{2} + a^2 -\sqrt{n + a^2} \ ,
\end{equation}
where $a$ is a dimensionless parameter defined by
$a \equiv \sqrt{ V / (8 k_d) }$, $V$ is the cell's volume and $k_d$ is the ratio of the dimerization
and undimerization rates.
If there are $n_{ 2 }(n)$ dimers in the cell, the equilibrium probability distribution  for the number
$j$ of dimers available through diffusion for binding to a 
promoter with characteristic volume $V_{ P }$ much smaller than the cell's volume $V$ is given by \cite{van1992stochastic}
\begin{equation}
\label{DimerPEq}
		{\cal P}_{ j }( \lambda n_{ 2 }( n ) )\ ,
\end{equation}
where ${\cal P}_{ j }( \theta )$
is the Poisson distribution of mean $\theta$ (evaluated at $j$), and $\lambda \equiv V_{ P } / V \ll 1$.
When writing down \eqref{DimerPEq} we have taken into account that for typical values of protein (and 
protein dimer) diffusion coefficients and dimerization rates, 
see~\cite{bialek2005physical,liu1993rapid,xu1998mechanism}, dimer formation and dissociation
within the small volume $V_P$ occurs with negligible probability compared to diffusion
into and out of $V_P$.
Note that  $\lambda$ is  typically very small, since promoters have linear dimensions in the nanometer 
range and cells in the micrometer range. As discussed for example in~\cite{benichou2009searching}, other transport mechanisms more 
efficient 
than three-dimensional diffusion must be at play that enable the promoter to gauge the actual number of 
molecules in the cell. Assuming that transport does not 
distinguish between dimers, and that the number of dimers does not influence the transport of a single 
dimer (essentially, that dimers are independent regarding transport, as is the case for diffusion), the 
distribution of dimers in $V_{ P }$ is binomial in general, with an ``effective rate of volumes" 
parameter $\lambda$. In the relevant limit $\lambda \ll 1$ we regain~\eqref{DimerPEq}.

We will now explicitly take into account that the promoter timescale is much shorter than the protein 
timescale by assuming that the distributions $p( P_{ f } , t )$,
$p( P_{ b } , t )$, have time to reach equilibrium for each fixed value of the number of proteins. 
Using the equilibrium dimer number distribution, we have:
\begin{equation}
\label{PSum}
	p^{ eq }( P ) =
		\sum_{ j \geqslant 0 }
			p^{ eq }( P \mid j )
			{\cal P}_{ j }( \lambda n_{ 2 } ( n ) ) \ ,
\end{equation}
with
$P \in \{ P_{ f } , P_{ b } \}$.
Solving equation~\eqref{PMeq} in equilibrium ($\dot{p} = 0$) and 
substituting in \eqref{PSum} leads to
\begin{subequations}
\label{PSol}
\begin{equation}
	\label{PSolf}
	p^{ eq }( P_{ f } \mid n ) =
		\sum_{ j \geqslant 0 }
			\frac{ 1 }
				{ 1 + k j }
			{\cal P}_{ j }( \lambda n_{ 2 }( n ) ) \ , 
\end{equation}
\begin{equation}
	\label{PSolb}
	p^{ eq }( P_{ b } \mid n ) =
		\sum_{ j \geqslant 0}
			\frac{ k j }
				{ 1 + k j }
			{\cal P}_{ j }( \lambda n_{ 2 }( n ) ) \ ,
\end{equation}
\end{subequations}
where we have introduced the dimensionless parameter $k \equiv k^{ + } / k^{ - }$ and we now emphasize the dependence on protein copy number $n$.


\subsection{mRNA and Protein Levels}
\label{mandp}

The production of mRNA and protein molecules in the cell has been found, in many cases, to occur in 
sharp geometrical  bursts~\cite{cai2006stochastic,kaufmann2007stochastic,ozbudak2002regulation,taniguchi2010quantifying}.
Although the concept of bursts
and the mechanisms underlying them are still open to discussion, see for 
example~\cite{paulsson2005models,elgart2011connecting}, a basic description stems from two
simple ideas.
First, if transcription/translation events are widely spaced compared to their duration, it is 
reasonable to speak of burst events.
Second, the geometric distribution relates to the number of consecutive ``heads" in the throwing of a 
(generally biased) coin; thus, if during a burst event there is a fixed probability that \emph{another} 
molecule will be produced, a geometrically distributed number of molecules results. A major achievement 
of this burst description is that the resulting predicted form of unregulated protein expression 
distributions \cite{friedman2006linking,shahrezaei2008analytical} is 
remarkably simple and fits an impressive number of experimental distributions measured for yeast 
populations ~\cite{taniguchi2010quantifying}. 
We adopt here an approach in which bursts are formulated in a stochastic framework 
both for transcription and for translation.

Owing to the timescale separation between promoter and mRNA/protein dynamics, for
transcriptional regulation the latter are described in chemical reaction notation by
\begin{equation}
\label{mpReac}
	\left\{
		\begin{aligned}
			\varnothing
				&\xrightarrow{ \beta_{ m } f( n )}
				&&\mu_{ m } m \ ,
			\\
			m
				& \;\;\; \xrightarrow{ \delta_{ m } }
				&&\varnothing \ ,
			\\
			m
				& \;\;\;\, \xrightarrow{ \beta_p }
				&&m + \mu_p p \ ,
			\\
			p 
				& \;\;\;\, \xrightarrow{ \delta_p }
				&&\varnothing \ .
		\end{aligned}
	\right .
\end{equation}
Here $m$ is the mRNA and $p$ is the protein, while $n$ stands for protein copy number. $f$ is the regulation function, such that:
\begin{equation}
\label{fDef}
\begin{aligned}
	f(n) = \sum_{ j \geqslant 0 }
				\frac{ 1 + \rho k j }
					{ 1 + k j }
				{\cal P}_{ j }( \lambda n_2( n ) ) \ .
\end{aligned}
\end{equation}
Thus, $\beta_{ m }$ is the transcription rate when the promoter is free, and $\rho \beta_{ m }$ is the transcription rate when the promoter is bound to a dimer; the protein exhibits negative auto-regulation (auto-inhibition) if $\rho < 1$, and positive auto-regulation (auto-activation) if $\rho > 1$;   $\mu_{ m }$ is the mean transcriptional burst size. With the burst scenario in mind, the transcription rates above are to be interpreted as the mean rates at which a transcription event takes place; this event is modeled as the instantaneous transcription of a certain number (drawn from a geometric distribution) of mRNA molecules. We assume here that regulation affects only the base transcription rate, and not burst size. Finally, $\delta_{ m }$ is the mRNA degradation rate.
Similar definitions stand for the protein parameters (with $\beta_p$ the translation rate, interpreted as the rate at which a single mRNA molecule initiates an instantaneous translational burst, $\mu_p$ the mean translational burst size, and $\delta_p$ the protein degradation rate).

\begin{figure}[h]
\includegraphics[scale=0.95]{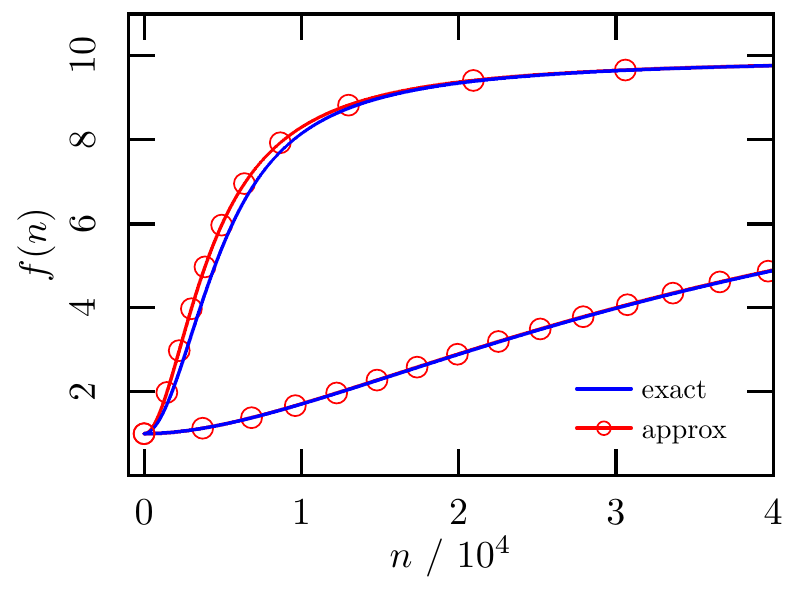}
\caption[Approximation for the regulation function]{(Color online) Aproximation \eqref{fApprox} for the regulation function. We fixed $\lambda = 10^{-2}$, $\rho = 10$, $a=10^2$ for a typical example.
\textbf{Bottom curves:} $k=10^{-2}$; \textbf{top curves:} $k=0.5$.
}
\label{Fi:regulation}
\end{figure}

It is interesting to see that the timescale separation for promoter dynamics allows all details of regulation to be condensed in the regulation function. Different regulatory dynamics affecting only the transcription rate and obeying the same timescale separation may be modeled in this framework simply by considering a different form of $f(n)$. Note also that a useful approximation to the regulation function as defined by \eqref{fDef} exists if $k \ll 1$. If $\lambda n_2( n )$ is small, the low $j$ terms of the sum will dominate; taylor expansion of the denominator to lowest order in $k j$ (for $k j \ll 1$) and explicit calculation of the sum leads to:
\begin{equation}
\label{fApprox}
	f( n ) \approx \frac{ 1 + \rho k \lambda n_2( n )  }{ 1 + k \lambda n_2( n ) } \ .
\end{equation}
If $\lambda n_2( n )$ is large, the large $j$ terms dominate, and the approximation given by \eqref{fApprox} remains valid because $( 1 + \rho k \alpha  ) / ( 1 + k \alpha ) \approx \rho$ for large $\alpha$. Direct numerical calculation reveals that \eqref{fApprox} is a good approximation overall, even for moderately large values of $k < 1$, see Figure~\ref{Fi:regulation}.

Let $E_i ( \theta ) \equiv  \frac{( \theta-1)^{ i - 1} }{ \theta ^i }$ be the geometric distribution of mean $\theta$ (evaluated at $i$), conditioned to non-zero values $i \geqslant 1$ because a burst of zero molecules has no physical meaning
\footnote{Note that, if $E^0(\theta)$ is the non-conditioned geometrical distribution of mean $\theta$, the relation $E_i( \theta ) = \frac{ \theta }{ \theta -1 } E^0_{ i } ( \theta -1 )$ holds for all $i \geqslant 1$. Thus, ``conditioned bursting" with frequency $\alpha$ and mean $\theta$ is equivalent to ``non-conditioned bursting" with frequency $\frac{ \theta }{ \theta - 1 } \alpha$ and mean $\theta -1$, and the difference becomes relevant only for $\theta \sim 1$.}.
Let also $p_{ j , n }( t )$ be the joint probability distribution of mRNA and protein  copy numbers (evaluated at mRNA copy number $j$ and protein copy number $n$) at time $t$. Then the Master Equation for the process \eqref{mpReac} reads
\begin{equation}
\label{2dMeq}
\begin{aligned}
	\dot{ p }_{ j , n }( t )& =
		\Big[
		\beta_m f( n )
		\sum_{ i \geqslant 1 }
			E_i( \mu_m )
			( \mathbb{ E }_m^{ -i } - 1 ) +
			\delta_m ( \mathbb{ E }_m - 1 ) j 
		\\
		&+ \ \beta_p j
		\sum_{ i \geqslant 1 }
			E_{ i }( \mu_p )
			( \mathbb{ E }^{ -i }_p - 1 ) +
			\delta_p ( \mathbb{ E }_p - 1 ) n
		\Big]
		 p_{ j , n }( t ),
\end{aligned}
\end{equation}
where we have made use of the ``step operators'' $\mathbb{ E }_{ m }$, $\mathbb{ E }_p$ defined by:
\begin{equation}
\label{StepOp}
\begin{aligned}
	\mathbb{ E }_m^i g_{ j , n } ( t ) &= g_{ j + i , n } ( t ) \ ,
	\\
	\mathbb{ E }^i_p g_{ j , n } ( t ) &= g_{ j , n + i } ( t ) \ ,
\end{aligned}
\end{equation}
for any function $g$ depending on mRNA copy number $j$, protein copy number $n$, and time $t$. With this notation, each term of the sum in the first term on the right hand side of \eqref{2dMeq} stands for the creation of $i$ mRNA molecules through a burst of size $i$, with bursts of arbitrary size occurring at rate $\beta_m f(n)$. The second term  represents the degradation of one mRNA molecule, occurring at rate $\delta_m j$. Each term of the sum in the third term on the right hand side  stands for the creation of $i$ protein molecules through a burst of size $i$, with bursts of arbitrary size occurring at rate $\beta_p j$. Finally, the last term in \eqref{2dMeq} represents the degradation of one protein molecule, occurring at rate $\delta_p n$. 

\subsection{Translational Regulation}
\label{Trans Level}

Consider now the case of translational regulation,  
Figure~\ref{Fi:scheme}, right arrow. In this case we assume that 
mRNA production proceeds through bursts without protein regulation, and that the rate of
production of protein bursts in translation is modulated by a regulation function $\tilde f( j, 
n ) $ depending on mRNA and protein copy numbers and describing an interaction (direct or indirect) of
the protein with its mRNA. Then, mRNA and protein dynamics
are described in chemical reaction notation by
\begin{equation}
\label{mpReacTrans}
	\left\{
		\begin{aligned}
			\varnothing
				& \;\;\; \xrightarrow{ \beta_{ m } }
				&&\mu_{ m } m \ ,
			\\
			m
				& \;\;\; \xrightarrow{ \delta_{ m } }
				&&\varnothing \ ,
			\\
			m
				& \;\;\;\, \xrightarrow{ \beta_p \tilde f( j, n ) }
				&&m + \mu_p p \ ,
			\\
			p 
				& \;\;\;\, \xrightarrow{ \delta_p }
				&&\varnothing \ .
		\end{aligned}
	\right .
\end{equation}

The Master Equation for the process \eqref{mpReacTrans} then reads
\begin{equation}
\label{2dMeqTrans}
\begin{aligned}
	\dot{ p }_{ j , n }( t ) =
		&\Big[
		\beta_m 
		\sum_{ i \geqslant 1 }
			E_i( \mu_m )
			( \mathbb{ E }_m^{ -i } - 1 ) +
			\delta_m ( \mathbb{ E }_m - 1 ) j 
		\\
		&+ \ \beta_p  j
		\sum_{ i \geqslant 1 }
			E_{ i }( \mu_p )
			( \mathbb{ E }^{ -i }_p - 1 ) \tilde f( j, n )
			\\
			 &+ \
			\delta_p ( \mathbb{ E }_p - 1 ) n
		\Big]
		 p_{ j , n }( t )\ .
\end{aligned}
\end{equation}


\section{Approximate Solutions for the Equilibrium Distributions}
\label{Solanal}
 
In what follows, $n$ and $j$ always stand for protein and mRNA copy numbers, respectively. 
The coupling between mRNA and protein reactions leads to correlations 
between the random variables corresponding to $n$ and $j$. 
As a result, the joint distribution $p_{ j , n }$ does not factorize and separate Master Equations for $n$ and $j$ do not exist. Studying the solutions of the Master Equations \eqref{2dMeq}, \eqref{2dMeqTrans} in general  calls for direct numerical simulations of the dynamics or 
numerical integration techniques. However, further timescale separations between mRNA and protein 
dynamics may be explored to simplify the problem.
We say mRNA is fast (compared to protein) if we can write the joint equilibrium distribution as:
\begin{equation}
\label{fastRNAseparation}
	p^{eq}_{ j , n } = q^{eq}_{ j | n } \, p^{eq}_n \ ,
\end{equation}
where $q^{eq}_{ j | n }$ is the equilibrium solution to the Master Equation for mRNA with fixed $n$. This means that mRNA dynamics are fast enough for a large number of mRNA-only reactions to take place 
before an $n$-changing reaction occurs, so that $q_{ j | n }$ reaches equilibrium and the time spent out of equilibrium is negligible. Then, by substituting \eqref{fastRNAseparation} in the appropriate 
general Master Equation and summing over $j$, we obtain an equation for $p^{ eq }_n$ independent of $j$. The physical idea is that for a certain $n$ the mRNA will essentially sample the distribution $q^{eq}_{ j | n }$, and $j$-dependent quantities are correspondingly averaged over this distribution.

Similarly, we say protein is fast if we may write
\begin{equation}
\label{fastproteinseparation}
	p^{eq}_{ j , n } = p^{eq}_{ n | j } \, q^{eq}_j \; ,
\end{equation}
with analogous interpretations. In this case, for each $j$ the $p^{eq}_{ n | j }$ distribution is sampled and $n$-dependent quantities are averaged over it.

We postpone to Section~\ref{Approx} the analysis of the conditions under which such a separation 
holds as a good approximation, and use it here to write down an equation for $p^{eq}_n$ or $q^{eq}_n$
from which approximate analytic expressions 
for the stationary solutions of the Master Equations \eqref{2dMeq}, \eqref{2dMeqTrans} will be derived. 

\subsection{Transcriptional Regulation Under Fast mRNA Dynamics}
\label{FastmRNA}

In this section we consider transcriptional regulation for the case of fast mRNA compared to protein dynamics. We explore both the discrete scenario and a continuous approximation.
It is convenient in this case to consider fixed $n$, since fast mRNA dynamics should 
allow mRNA copy number 
to equilibrate for each fixed protein copy number. This means we are considering the reactions
\begin{equation}
\label{mReac}
\left\{
	\begin{aligned}
		\varnothing
			&\xrightarrow{ \beta_{ m } f( n )}
			&&\mu_{ m } m \ ,
		\\
		m
			& \;\;\; \xrightarrow{ \delta_{ m } }
			&&\varnothing \ ,
	\end{aligned}
	\right .
\end{equation}
at fixed $n$. Let $q_{ j \mid n } ( t )$ be the distribution of mRNA copy number (evaluated at $j$) at 
time $t$, given $n$. The Master Equation for this process has the simple 
form:
\begin{equation}
\label{fastmmCondMeq}
\begin{aligned}
	\dot{ q }_{ j \mid n }( t ) =
		&\Big[
		\beta_m f( n )
		\sum_{ i \geqslant 1 }
			E_{ i }( \mu_m )
			( \mathbb{ E }_m^{ -i } - 1 ) 
		\\
		&+ \
			\delta_m ( \mathbb{ E }_m - 1 ) j
		\Big]
		q_{ j \mid n }( t ) \ .
\end{aligned}
\end{equation}

Let  $q^{ eq }_{ \mid n }$ be the equilibrium distribution of mRNA copy number, for each protein copy number $n$. 
The mean value of mRNA corresponding to this distribution can be found to be (see Appendix~\ref{app:MeanmRNA})
\begin{equation}
\label{fastmmCondSol}
	\langle id \rangle_{ q^{ eq }_{ \mid n } } =
		\mu_{ m } \gamma_{ m } f( n ) \ ,
\end{equation}
where $\gamma_{ m } = \beta_{ m } /  \delta_{ m }$, and $id$ is the identity function.


In the protein timescale, we have the reactions:
\begin{equation}
\label{pReac}
	\left\{
		\begin{aligned}
			m
				& \;\;\, \xrightarrow{ \beta_p }
				&&m + \mu_p p \ ,
			\\
			p 
				& \;\;\, \xrightarrow{ \delta_p }
				&&\varnothing \ .
		\end{aligned}
	\right .
\end{equation}
Let $p_n( t )$ be the distribution of protein copy number (evaluated at $n$) at time $t$.
According to \eqref{fastRNAseparation}, the 
Master Equation for this process reads
\begin{equation}
\label{fastmpMeq}
\begin{aligned}
	\dot{ p }_{n}( t ) \! &= \!
		 \Big[
			\sum_{\substack{j \geqslant 0,\\ i \geqslant 1}}
					\! \beta_p
					E_{i}(\mu_p)
				 	(\mathbb{E}^{-i}_p\! -1)
					j q^{eq}_{j \mid n } \! +
					\delta_p( \mathbb{E}_p\!-1) n
		\Big]
		p_{ n }( t ) 
	\\
	&=
		\Big[
			r \delta_p \!
			\sum_{ i \geqslant 1 }
			E_{ i }( \mu_p )
				( \mathbb{ E }_p^{ -i }\!-1) f( n ) \! +
		\delta_p( \mathbb{ E }_p\!-1 ) n
		\Big]
		p_{ n }( t ),
\end{aligned}
\end{equation}
where $r \equiv  \mu_{ m } \gamma_{ m } \gamma_{p}$ and $\gamma_{p} \equiv \beta _p/ \delta _p$. 
The parameter $r$ is the prefactor of the average effective rate of translation
burst events  scaled by the degradation rate of the protein, $r f(n) $. We will see that,
together with the average translational burst size $\mu _p$, it determines the protein equilibrium distribution in this approximation.

As expected, when mRNA is fast, protein dynamics depends at each time only on the average mRNA corresponding to the available protein number $n$. Specifically, the translation rate becomes proportional to $\langle id \rangle_{ q^{ eq }_{ \mid n } }$, which is in turn proportional to $f(n)$. Through this mechanism, promoter-level regulation gauges the number of proteins  present in the cell at a certain time. Note also that further details of mRNA dynamics, including burst-like production, are lost at the level of protein.

Let us consider as well a continuous approximation of the dynamics. For this we take $x \equiv \lambda n$ as an ``approximately continuous" variable (recall that $\lambda \ll 1$). A continuous Master Equation for the distribution $p( x , t )$ of protein ``concentration" $x$ reads (see Appendix~\ref{app:ContApprox})
\begin{equation}
\label{fastmpContMeq}
\begin{aligned}
	\dot{p}(x, t) =
		& r \delta_p \! 
		\int_{0}^{x}
			\! \! \! f(y)
			\Big[
				E(x - y, \tilde\mu_p) - \delta_D (x - y)
			\Big] p(y, t) dy 
		\\
		&+ \delta_p \, \partial_{x}
			\Big[
				x p(x, t)
			\Big] \ ,
\end{aligned}
\end{equation}
where $E( x , \theta ) \equiv ( 1 /Ê\theta ) e^{ -x / \theta }$ is the exponential probability distribution of mean $\theta$ 
evaluated at $x$ and $\delta_D$ is the Dirac delta. For simplicity we have chosen to keep the symbol $f$, such that $f( x ) = f( n )$ 
for $x = \lambda n$. The exponential distribution term accounts for the contribution to $p ( x, t )$ due to bursts leading to 
concentration $x$, and the Dirac delta term accounts for bursts away from $x$; $\tilde\mu_p$ is the
 rescaled burst size, 
$\tilde\mu_p\equiv \lambda \mu_p$. The last term is due to protein degradation.


The equilibrium solution of \eqref{fastmpContMeq} can be found to be 
(see Appendix~\ref{app:ContSol})
\begin{equation}
\label{fastmpContSol}
	p^{ eq }( x ) =
		A_c \, x^{ -1 } e^{ - x / \tilde\mu_p}
		e^{ r \int_c^{ x } du f( u ) / u } \ ,
\end{equation}
where the constant $A_c$ depends on the arbitrary constant $c$ and is determined by normalization.

If we solve equation \eqref{fastmpMeq} directly in the discrete setting (see Appendix~\ref{app:Sol}), we find the solution:
\begin{equation}
\label{fastmpSol}
	p_n^{ eq } =
	\frac{ r \, p_0^{ eq } }{ n }
	\prod_{ i = 1 }^{ n - 1 }
	\left(
			r \frac{ f( i ) }{ i } + \frac{ \mu_p - 1 }{ \mu_p }
	\right) \ ,
\end{equation}
for $n \geqslant 1$, with $p_0^{ eq }$ determined by normalization.

Generically, the continuous approximations presented throughout this section are very accurate for burst sizes of order 10 and higher. It should be noted, however, that very sharp peaks (with a width of the order of a single molecule) that arise for zero protein or mRNA in some parameter ranges are not well captured by the continuous approximation.

The role of the biological parameters in the qualitative features of the protein distribution is particularly clear in the continuous setting. 
To study some of these features, consider the derivative of the probability distribution given by~\eqref{fastmpContSol}; concentrations $x > 0$ 
where probability peaks correspond to $\partial_x p^{eq}(x)=0$, leading to
\begin{equation}
\label{peaks}
	r \tilde\mu_p f(x) = x + \tilde\mu_p \ .
\end{equation}
Let us consider the regulation function as given by the approximation described by~\eqref{fApprox}. In the continuous description we write:
\begin{equation}
	f( x ) \approx \frac{ 1 + \rho k x_2( x )  }{ 1 + k x_2( x ) } \ ,
\end{equation}
with
\begin{equation}
	x_2(x) \equiv \lambda n_2(n) = \frac{x}{2} + \tilde a^2 -\sqrt{x + \tilde a^2} \  
\end{equation}
and $\tilde a = a / \sqrt{\lambda}$. By noting that equation~\eqref{peaks} is equivalent to a quartic equation in $z = \sqrt{x + \tilde a^2}$, it is easy to prove that $p^{eq}$ is at most bimodal (see Appendix~\ref{app:Quartic}).

\begin{figure}[h]
\includegraphics[scale=0.95]{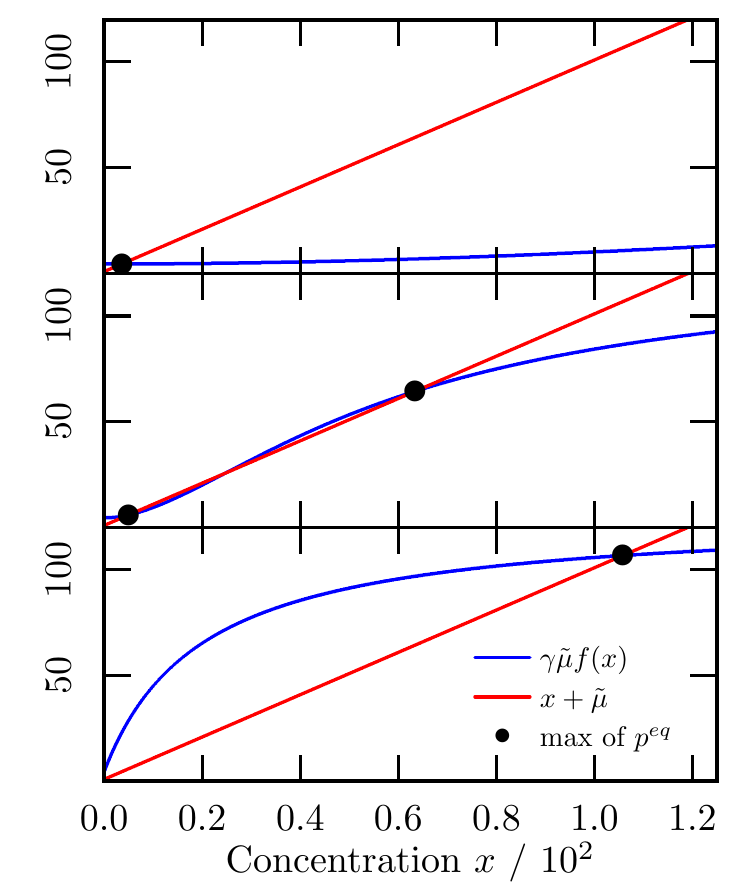}
\caption[Role of dimerization in protein equilibrium distributions]{(Color online) Illustration of the effect of varying the dimerization parameter $\tilde a$ when bimodality is possible. For low dimerization (\textbf{top}) there is only a low concentration equilibrium, and for high dimerization (\textbf{bottom}) there is only a high concentration equilibrium. Bimodality without a peak at zero arises only for intermediate dimerization (\textbf{middle}). Example parameters are $r = 5$, $\tilde \mu _p= 0.9$, $\rho = 28$, $k=10^{-1}$, and:
\textbf{Top:} $\tilde a = 50$; \textbf{middle:} $\tilde a = 5$; \textbf{bottom:} $\tilde a = 0$.
}
\label{Fi:dimers}
\end{figure}

\begin{figure}[h]
\includegraphics[scale=0.95]{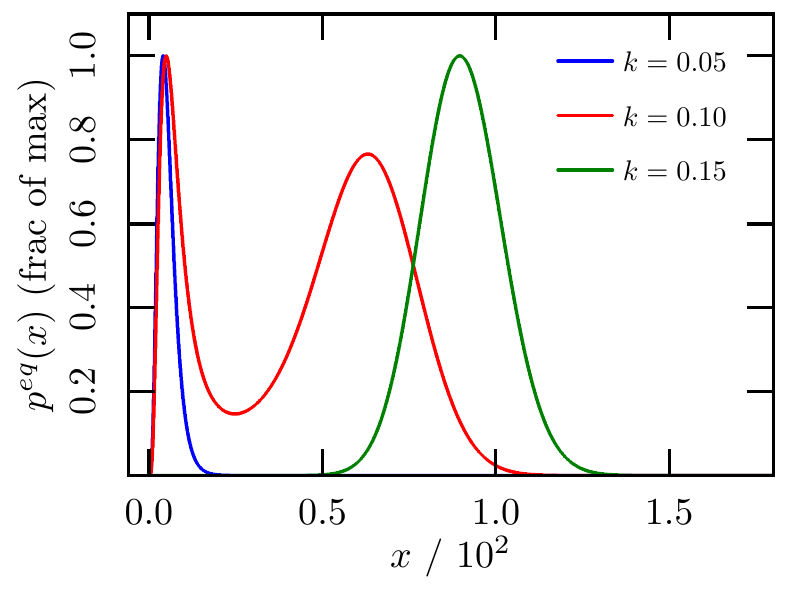}
\caption[Role of promoter affinity in protein equilibrium distributions]{(Color online) Illustration of the effect of varying promoter affinity $k$ when bimodality is possible. We fixed $r = 5$, $\tilde\mu_p = 0.9$, $\rho = 28$ and $\tilde a=5$.}
\label{Fi:affinity}
\end{figure}

In the case of negative auto-regulation ($\rho < 1$), $p^{eq}$ is always unimodal because the regulation function
is monotonically decreasing. Positive auto-regulation ($\rho~>~1$) is necessary for more structured distributions, and bimodal 
distributions do in fact arise for some parameter sets. It is interesting to note that in the limit of weak dimerization (large $\tilde a$) 
$p^{eq}$ is always unimodal, while in the limit of strong dimerization (small $\tilde a$) it is unimodal if $\gamma > 1$ and bimodal with 
a peak at zero if $\gamma < 1$; bimodal distributions that do not peak at zero are present only for intermediate dimerization 
(see Figure~\ref{Fi:dimers}). Near parameter regions allowing for bimodality, 
the shape of $p^{eq}$  is also very sensitive to the
promoter affinity $k$ (see Figure~\ref{Fi:affinity}). Varying $r$ and $\rho$ 
affects bimodality as well, but the  values of these parameters have a  stronger effect on the peak positions.  Finally, the burst size parameter $\tilde\mu_p$ also affects the position and relative size of peaks in $p^{eq}$, but its essential role is to produce the heavy tailed distributions commonly observed experimentally.

It is now easy to obtain the distribution of mRNA expression. For 
the continuous approximation, taking into account the Master Equation \eqref{fastmmCondMeq}, we have as in \eqref{fastmpContMeq}
\begin{equation}
\label{fastmmContCondMeq}
\begin{aligned}
	&\dot{ q }( z , t \mid x ) = \delta_m \, \partial_{ z }
			\Big[
				z q( z , t \mid x )
			\Big] 
		\\
		&+ \;
	    \beta_{ m } f( x )
		\int_{ 0 }^{ z }
			\Big[
				E( z - w  , \tilde\mu_m ) - \delta( z - w )
			\Big] q( w , t \mid x ) \, dw .
\end{aligned}
\end{equation}
This is an evolution equation for the distribution of a ``continuous" mRNA concentration variable $z \equiv \lambda j$, given a fixed protein concentration $x = \lambda n$ (with $\tilde\mu_m$ again a rescaled burst size). Since $f$ depends on protein but not mRNA concentration, we find for the equilibrium distribution (see Appendix~\ref{app:ContSol}) a Gamma distribution:
\begin{equation}
\label{fastmmContCondSol}
	q^{ eq }( z \mid x ) = G( z , \gamma_{ m } f( x ), \tilde\mu_m ) \ .
\end{equation}
To find the equilibrium distribution of mRNA, we take the integral over all values of protein concentration, weighted by the respective probabilities given by \eqref{fastmpContSol}:
\begin{equation}
\label{fastmmContSol}
\begin{aligned}
	q^{ eq }( z ) &=
		\int_{ 0 }^{ \infty } q^{ eq }( z \mid x ) p^{ eq }( x ) \, dx
	\\
	&=
		\left\langle
			G( z , \gamma_{ m } f, \tilde\mu_m )
		\right\rangle_{ p^{ eq } }  \ .
\end{aligned}
\end{equation}

Similarly, the solution for the discrete dynamics, corresponding to equation~\eqref{fastmmCondMeq}, is given by a Negative Binomial distribution (c.f.~Appendix~\ref{app:Sol}):
\begin{equation}
	q^{ eq }_{ j \mid n } =
	N_j \left(
		 \frac{ \mu_m }{ \mu_m  - 1 } \gamma_m f( n ), \frac{ 1 }{ \mu_m }
	\right) \ .
\end{equation}
The discrete equilibrium distribution for mRNA is found in this case by summing over all $n$, weighing with the discrete protein distribution given by \eqref{fastmpSol}:
\begin{equation}
\label{fastmmSol}
\begin{aligned}
	q^{ eq }_j &=
		\sum_{ n \geqslant 0 } q^{ eq }_{ j \mid n } p^{ eq }_n \, 
	\\
	&=
		\left\langle
			N_j \left(
				\frac{ \mu_m }{ \mu_m  - 1 } \gamma_m f, \frac{ 1 }{ \mu_m }
			\right)
		\right\rangle_{ p^{ eq } } \ .
\end{aligned}
\end{equation}
The performance of the continuous approximation is similar for mRNA and for
protein.


\subsection{Transcriptional Regulation Under Fast Protein Dynamics}
\label{FastProtein}


It is now convenient to consider fixed $j$, since in this case protein dynamics is much faster and will equilibrate. Let $p_{ n \mid j } ( t )$ be 
the distribution of protein copy number (evaluated at $n$) at time $t$, given $j$. We have again reactions \eqref{pReac}, but in this case 
we write the Master Equation for fixed mRNA copy number $j$:
\begin{equation}
\label{fastppCondMeq}
	\dot{ p }_{ n \mid j }( t ) = \!
		\Big[
		\beta_p j
		\sum_{ i \geqslant 1 }
			E_{ i }( \mu_p )
			( \mathbb{ E }_p^{ -i } \!- \!1 ) +
			\delta_p ( \mathbb{ E }_p \!- \!1 ) n
		\Big]
		p_{ n \mid j }( t ) .
\end{equation}

In the continuous approximation, we find:
\begin{equation}
\label{fastppContCondMeq}
\begin{aligned}
	\dot{ p }( x , t \mid z ) &= \tilde\gamma_p \delta_p
		 \int_{ 0 }^{ x } \!
			\Big[
				E(x - y, \tilde\mu_p ) - \delta_D (x - y)
			\Big] p(y, t) \, dy 
	\\
		& \quad +  \delta_p \, \partial_{ x }
			\Big[
				x p( x , t )
			\Big] \ ,
\end{aligned}
\end{equation}
where
$\tilde\gamma_p \equiv \gamma_p / \lambda $.
This equation can be solved for the equilibrium distribution in exactly the same way as equation~\eqref{fastmmContCondMeq}, yielding:
\begin{equation}
\label{fastppContCondSol}
	p^{ eq }( x \mid z ) =
		G( x , \tilde\gamma_p z, \tilde\mu_p ) \ .
\end{equation}

Similarly, the discrete solution (equation~\eqref{fastppCondMeq}) is:
\begin{equation}
\label{fastppCondSol}
	p^{ eq }_{ n \mid j } =
		N_n \left(
			 \frac{ \mu _p} {\mu _p - 1 } \gamma_p j, \frac{ 1 }{ \mu _p}
		\right) \;,
\end{equation}
where
$N_n(0 , \cdot ) \equiv \delta_{ n , 0 }$,
with $\delta_{ n , 0 }$ a Kronecker delta symbol.


Following arguments similar to those leading to equation~\eqref{fastmpMeq}, the Master Equation for mRNA reads in this case:
\begin{equation}
\label{fastpmMeq}
\begin{aligned}
	\dot{ q }_{ j }( t ) &=
		\Big[
			\beta_m
			\sum_{ i \geqslant 1 }
			E_{ i }( \mu_m )
				 	( \mathbb{ E }_m^{ -i } - 1 )
					\langle f \rangle_{ p^{ eq } _{ \mid j } } 
					\\
					&+ \;
					\delta_m ( \mathbb{ E }_m - 1 ) j
		\Big]
		q_{ j }( t ) \ ,
\end{aligned}
\end{equation}
and the corresponding continuous Master Equation is
\begin{equation}
\label{fastpmContMeq}
\begin{aligned}
	 & \dot{ q }( z , t )= \delta_m \, \partial_{ z }
			\Big[
				z q( z , t )
			\Big] \  +
	\\
		& + \beta_m \! \int_{ 0 }^{ z }
			\!\! \langle f \rangle_{ p^{ eq }( \mid w ) }
			\Big[
				E(z \! - \! w, \tilde\mu_m ) \! - \! \delta_D (z-w)
			\Big] q(w, t) dw .
\end{aligned}
\end{equation}


The equilibrium solution of equation~\eqref{fastpmContMeq} can be found through the same method as the one used for equation~\eqref{fastmpContMeq}, yielding:
\begin{equation}
\label{fastpmContSol}
	q^{ eq }( z ) =
		A_c z^{ -1 } e^{ - z / \tilde\mu_m }
		e^{ \gamma_m
			\int_c^{ z } du
				\langle f \rangle_{ p^{ eq }( \mid u ) } / u } \ ,
\end{equation}
where $A_c$ is again a normalization constant.
The discrete solution, for equation~\eqref{fastpmMeq}, is
\begin{equation}
\label{fastpmSol}
	q_j^{ eq } =
	\frac{ \gamma_m \,
	q_0^{ eq } }{ j }
	\prod_{ i = 1 }^{ j - 1 }
	\left(
			\gamma_m \frac{ \langle f \rangle_{ p^{ eq }_{ \mid i } }  }{ i } +
			\frac{ \mu_m - 1}{ \mu_m }
	\right) \ ,
\end{equation}
for $j \geqslant 1$, with $q_0^{ eq }$ determined by normalization (note that $\langle f \rangle_{ p^{ eq }_{ \mid 0 } } =
f( 0 ) = 1$).


In the continuous approximation, the distribution of protein concentration follows immediately from the integration of the conditional distribution given by equation~\eqref{fastppContCondSol}:
\begin{equation}
\label{fastppContSol}
\begin{aligned}
	p^{ eq }( x ) &=
		\int_{ 0 }^{ \infty } p^{ eq }( x \mid z ) q^{ eq }( z ) \, dz
	\\
	&=
		\left\langle
			G( x , \tilde\gamma_p \, id, \tilde\mu_p )
		\right\rangle_{ q^{ eq } } \ .
\end{aligned}
\end{equation}

The corresponding discrete distribution is:
\begin{equation}
\label{fastppSol}
\begin{aligned}
	p^{ eq }_n &=
		\sum_{ j \geqslant 0 } p^{ eq }_{ n \mid j } q^{ eq }_j
	\\
	&=
		\left\langle
			N_n \left(
				 \frac{ \mu_p }{ \mu_p  - 1 } \gamma_p \, id, \frac{ 1 }{ \mu_p }
			\right)
		\right\rangle_{ q^{ eq } } \ .
\end{aligned}
\end{equation}

As expected, in this timescale regime the role of the regulation function is confined to the level of mRNA. The protein distribution depends only on the mRNA distribution, plus translation rate and protein burst size. 


\subsection{Translational Regulation}  
\label{translational}

In this scenario, mRNA production takes place through bursts without protein regulation and so
mRNA reaches equilibrium independently of protein concentrations. Formally, mRNA dynamics
decouples from the general
Master Equation \eqref{2dMeqTrans}, yielding for the mRNA distribution $q_{ j }( t ) $ the Master Equation:
\begin{equation}
	\dot{ q }_{ j }( t ) =
		\Big[
		\beta_m
		\sum_{ i \geqslant 1 }
			E_{ i }( \mu_m )
			( \mathbb{ E }_m^{ -i } - 1 ) +
			\delta_m ( \mathbb{ E }_m - 1 ) j
		\Big]
		q_{ j }( t ) .
\end{equation}
The equilibrium solution for an unregulated process of this type, see Appendix~\ref{app:Sol}, is a Negative Binomial:
\begin{equation}
\label{m_unregulated}
	q^{eq}_j = N_j \left(
			\frac{  \mu_m }{ \mu_m - 1 } \gamma_m , \frac{ 1 }{ \mu_m }
		\right) \ ,
\end{equation}
whose average is $\gamma_m \mu_m$. 

In the fast mRNA dynamics approximation, the Master Equation for protein abundances reads
\begin{equation}
\label{fastmpMeqTrans}
\begin{aligned}
	\dot{ p }_{n}( t ) \! & = 
		\Big[
			\sum_{ i \geqslant 1 }
			\beta_p E_{ i }( \mu_p )
				( \mathbb{ E }_p^{ -i }\!-1) \langle id \tilde f( \cdot , n ) \rangle_{ q^{ eq }} 
				\\& + 				
		\delta_p( \mathbb{ E }_p\!-1 ) n
		\Big]
		p_{ n }( t ),
\end{aligned}
\end{equation}
which, with the simple regulation function $\tilde f( j , n ) = f(n)$, reduces to
\begin{equation}
\label{fastmpMeqTrans0}
	\dot{ p }_{n}( t ) =
		\Big[
		r \delta_p
		\sum_{ i \geqslant 1 }
			E_{ i }( \mu_p )
			( \mathbb{ E }_p^{ -i } - 1 ) f( n ) + \delta_p
			( \mathbb{ E }_p - 1 ) n
		\Big]
		p^{ eq }_n \ .
\end{equation}
For a general regulation function $\tilde f( j , n )$, \eqref{fastmpMeqTrans0} still
holds, where now
$f(n) \equiv \langle id \tilde f( \cdot, n ) \rangle_{q^{eq}}/(\gamma_m \mu_m)$. 

Equation~\eqref{fastmpMeqTrans0} is the same equation that 
describes the distribution of protein with transcriptional regulation
under fast mRNA dynamics (compare to 
equation~\eqref{fastmpMeq}). We thus see that the protein equilibrium distribution is the 
same that was found in Subsection \ref{FastmRNA}, with the appropriate interpretation
of the new regulation function $f$. Moreover, as we will see in 
Section~\ref{Approx}, this solution holds under less stringent conditions than that of
equation~\eqref{fastmpMeq}.

Finally let us consider the fast protein approximation in the translational regulation
scenario. By the same arguments of Subsection \ref{FastProtein}, the equilibrium protein
distribution will be given by 
\begin{equation}
\label{fastppTlSol}
	p^{ eq }_n =
		\sum_{ j \geqslant 0 } p^{ eq }_{ n \mid j } q^{ eq }_j \ ,
\end{equation}
where  $q^{ eq }_j$ is the Negative Binomial \eqref{m_unregulated} and $p^{ eq }_{ n \mid j }$ is the
equilibrium distribution for fixed mRNA copy number $j$. The latter is the stationary solution to
\begin{equation}
\label{fastppMeqTrans}
\begin{aligned}
	\dot{ p }_{n\mid j}( t ) \! = \!
		 &\Big[
			\sum_{ i \geqslant 1}
					\beta_p
					E_{i}(\mu_p)
				 	(\mathbb{E}^{-i}\! -1)
					j \tilde f( j , n )
\\					
					& +
					\delta_p( \mathbb{E}_p\!-1) n
		\Big]
		p_{ n \mid j}( t ) \ ,
\end{aligned}
\end{equation}
already found in Subsection \ref{FastmRNA} (see \eqref{fastmpMeq}, \eqref{fastmpSol}) to be given by
\begin{equation}
\label{fastppSolTrans}
	p_{n \mid j}^{ eq } =
	\frac{ j \gamma_p \, p_0^{ eq } }{ n }
	\prod_{ i = 1 }^{ n - 1 }
	\left(
			j \gamma_p \frac{ \tilde f(j, i ) }{ i } + \frac{ \mu_p - 1 }{ \mu_p }
	\right) \ ,
\end{equation}
or, in the continuous approximation for $n$ (see \eqref{fastmpContSol}), by
\begin{equation}
\label{fastmpContSolTrans}
	p^{ eq }_{\mid j}( x ) =
		A_c \, x^{ -1 } e^{ - x / \tilde\mu_p}
		e^{ j \gamma_p \int_c^{ x } du \tilde f(j, u ) / u } \ .
\end{equation}

For the purpose of comparison of this approximation with simulational results, 
we will use the analytic solution given by \eqref{fastppTlSol} with \eqref{m_unregulated} and the continuous approximation for $p^{ eq }_{ n \mid j }$  given by \eqref{fastmpContSolTrans}.

 
\section{Validity of the timescale separation approximations}
\label{Approx}

In this section we study the conditions under which the timescale separation assumptions
used in Section  \ref{Solanal} should hold approximately.
We illustrate the quality of the
approximate analytic solutions for protein
by comparing them with the results of simulations of the full
stochastic process described by the Master Equations \eqref{2dMeq}, \eqref{2dMeqTrans} using the Gillespie algorithm \cite{gillespie2001approximate}.
In order to illustrate the agreement with the analytic distributions, the simulation curves shown below were plotted for sampling sizes such that the error bars at each data point are smaller than the markers. We have checked that the structure of the curves obtained from these simulations is robust down to order $10^5$ independent samples.

Let the subscripts $f$ and $s$ refer to the fast and slow species respectively, and let $\alpha$ and $\sigma$ be the mean and standard deviation of the equilibrium distribution, respectively. Consider also the average times $T$ for a change of one molecule to occur. Two conditions must be met:

\begin{enumerate}
\item The fast species must approach equilibrium quickly compared to $T_s$. If a change 
of the slow 
species produces a change of absolute value $\Delta \alpha_{f }$ in the equilibrium average of the fast species, we 
must have:
\begin{equation}
\frac{ T_f^{up}}{ T_s } \Delta \alpha_{f} \ll 1 \  ,
\end{equation}
where $T_f^{up}$ is the average time for the production of one copy of the fast species, since it is straightforward to check for each case that re-equilibrating following a burst is the most demanding scenario. 
\item The fast species must accurately sample the equilibrium average within a time interval $T_s$.  The 
relative standard error of the mean for $N$ \emph{uncorrelated} samples from a distribution of mean 
$\alpha$ and standard deviation $\sigma$ is given by:
\begin{equation}
\epsilon = \frac{ \sigma }{ \alpha \sqrt{N} } \  .
\end{equation}
When the fast species dynamically samples the equilibrium distribution, two uncorrelated measurements 
will be spaced in time approximately by the correlation time $\tau_f = 1/\delta_f$. Thus, considering 
the relative error in an interval $T_s$, we find the condition:
\begin{equation}
\frac{ \sigma_f }{ \alpha_f \sqrt{T_s \delta_f} } \ll 1 \ . 
\end{equation}
\end{enumerate}

We now study the constraints imposed by applying conditions 1.\ and 2.\ self-consistently in the hypothesis of fast mRNA and fast protein, with both transcriptional and translational regulation.
For transcriptional regulation under fast mRNA we have
\begin{equation}
\begin{aligned}
	\alpha_f &= \gamma_m f(n) \mu_m \ ,\\ 
	\sigma_f &= \sqrt{ \gamma_m f(n) } \mu_m \ .
\end{aligned}
\end{equation}
After a protein event leading to $n$ we may write:
\begin{equation}
\begin{aligned}
	T_f^{up} &= \Big[ \mu_m \beta_m f(n) \Big]^{-1} \ ,\\
	\Delta \alpha_{f } &= \mu_m \gamma_m \Delta f \ ,
\end{aligned}
\end{equation}
where $\Delta f$ is the absolute value of the change in the value of $f$ associated with the protein event.
Since production and degradation reactions must be balanced in equilibrium, we may estimate 
$T_s$ for macroscopically occupied $n$ as:
\begin{equation}
	T_s = [ \mu_p \beta_p \mu_m \gamma_m f( n ) ]^{-1} \ ,
\end{equation}
Then, setting $\delta = \delta_p/\delta_m$ we find 
for condition 1.\ 
\begin{equation}
	\delta \mu_p r \Delta f \ll 1\ ,
\end{equation}
and for condition 2.\ 
\begin{equation}
	\sqrt{ \delta \mu_p r/ \gamma_m } \ll 1 \ .
\end{equation}
We may combine the two conditions and write:
\begin{equation}
\label{condTcfastm}
	\sqrt{ \delta r \mu_p }
	\left(
		\sqrt{ \delta r  \mu_p } \Delta f + 1 /  \sqrt{ \gamma_m }
	\right)
	\ll 1 \ .
\end{equation}
Note that $\Delta f$ is bounded  by $| \rho - 1 |$.
It should be mentioned that in the interpretation of a protein translation burst
as the protein production of a single mRNA molecule along its whole lifetime,
$\beta_p = \delta_m  (\mu_p - 1) / \mu_p$ (see [33]), $\delta r \mu_p = ( \mu_p - 1 ) \mu_m \gamma_m $ and 
the conditions of the fast mRNA approximation may be met only for very low protein burst sizes, even in the absence
of regulation.

\begin{figure}[h]
\includegraphics[scale=0.95]{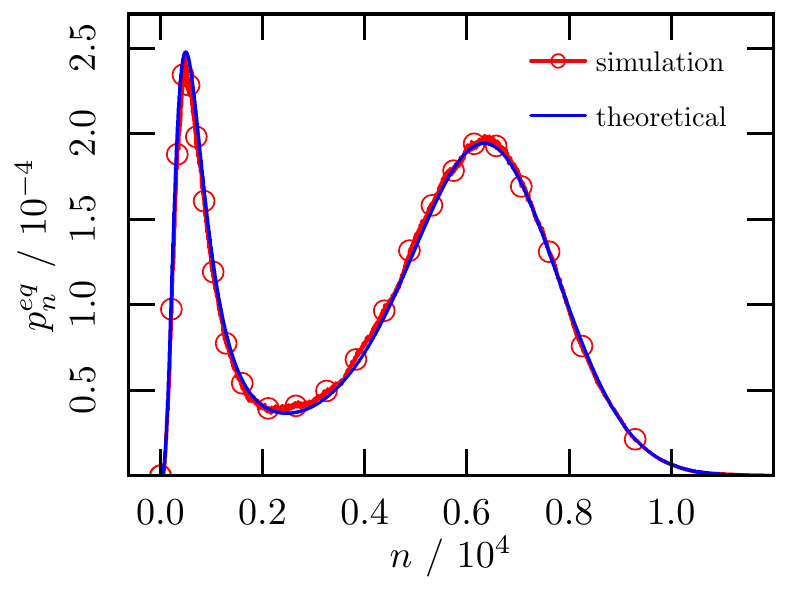}
\caption[]{(Color online) Illustration of the fast mRNA approximation with transcriptional regulation. Parameters are
$r = 5$, $\mu_p = 90$,
$\gamma_m = 2.25\e2$, $\mu_m = 2$, $\delta = 10^{-2}$,
$\rho = 28$, $k = 0.1$, $a = 50$ and $\lambda = 10^{-2}$.
Error bars are smaller than markers.}
\label{Fi:TSTcfastm}
\end{figure}

In Figure~\ref{Fi:TSTcfastm} we plot the equilibrium protein distribution obtained from simulations of the stochastic process \eqref{2dMeq}, together with the analytic solution
\eqref{fastmpSol}. The parameter values were chosen taking into account condition 
\eqref{condTcfastm}. There is excellent quantitative agreement between approximate and exact solutions.

\begin{figure}[h]
\includegraphics[scale=0.95]{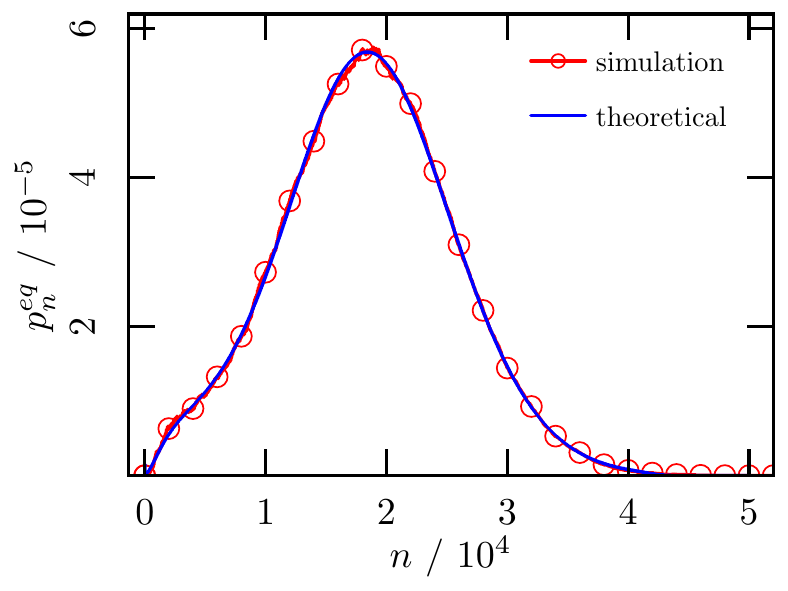}
\caption[]{(Color online) Illustration of the fast protein approximation with transcriptional regulation. Parameters are
$\gamma_p = 3$, $\mu_p = 20$,
$\gamma_m = 3$, $\mu_m = 20$, $\delta = 10^2$,
$\rho = 7.5$, $k = 0.25$, $a = 200$ and $\lambda = 10^{-2}$.
Error bars are smaller than markers.}
\label{Fi:TSTcfastp}
\end{figure}

For transcriptional regulation under fast protein we have, after an mRNA event leading to $j$:
\begin{equation}
\begin{aligned}
	\alpha_f &= \mu_p \gamma_p j \ ,\\ 
	\sigma_f &= \sqrt{ \gamma_p j } \mu_p \ ,\\
	T_f^{up} &= [\mu_p \beta_p j ]^{-1} \ ,\\
	T_s &= [ \delta_m j ]^{-1} \ .\\
\end{aligned}
\end{equation}
Since the variation of $j$ due to a burst is of order $\mu_m$, this leads to:
\begin{equation}
	\mu_m / \delta \ll 1
\end{equation}
for condition 1., and for condition 2.\ we find:
\begin{equation}
	1 / \sqrt{ \delta \gamma_p } \ll 1 \ .
\end{equation}
The combined condition is:
\begin{equation}
\label{condTcfastp}
	\frac{ 1 }{ \sqrt{ \delta } }
	\left(
		\mu_m / \sqrt{ \delta } + 1 / \sqrt{ \gamma_p }
	\right) \ll 1 \ .
\end{equation}

In Figure~\ref{Fi:TSTcfastp} we illustrate the behavior of the analytic solution
given by \eqref{fastppSol} versus simulations of the full stochastic process \eqref{2dMeq}. The parameter values were chosen taking into account condition 
\eqref{condTcfastp}, and once again there is excellent quantitative agreement.

\begin{figure}[h]
\includegraphics[scale=0.95]{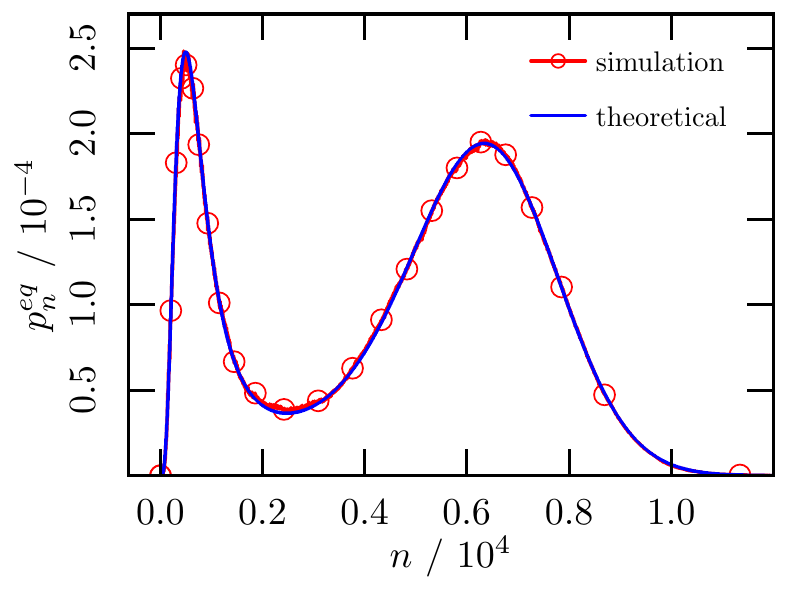}
\caption[]{(Color online) Illustration of the fast mRNA approximation with translational regulation. We took $\tilde f(j,n) = f(n)$, and parameters are
$r = 5$, $\mu_p = 90$,
$\gamma_m = 6.3\e4$, $\mu_m = 2$, $\delta = 1$,
$\rho = 28$, $k = 0.1$, $a = 50$ and $\lambda = 10^{-2}$.
Error bars are smaller than markers.}
\label{Fi:TSTlfastm}
\end{figure}

Consider now the case of translational regulation. The fast mRNA approximation
 may be treated in much the same way as the corresponding transcriptional regulation case. Note however that the mRNA-only reactions now decouple, and the equilibrium solution for the mRNA distribution does not depend on $n$. Thus $\Delta  \alpha_{ f } = 0$, and condition 1.\ imposes no constraints. The resulting constraint is due to condition 2.\ only and becomes:
\begin{equation}
	\sqrt{ \frac{ \delta r \mu_p f( n ) }{ \gamma_m }} \ll 1 \ ,
\end{equation}
to be considered for macroscopically-occupied values of $n$. Recall that $f(n)$ is bounded by $\text{max}(1,\rho)$.
Figure~\ref{Fi:TSTlfastm} again shows excellent quantitative agreement between
the analytic solution derived in Subsection  \ref{translational} for fast mRNA dynamics
and the equilibrium distributions obtained from simulations of \eqref{2dMeqTrans}. 

Notice the similarity between Figure~\ref{Fi:TSTlfastm} and Figure~\ref{Fi:TSTcfastm},
which is due to the fact that we have considered for the translational regulation
function $\tilde f(j, n )= f(n)$, with $f(n)$ the transcriptional regulation function used for the results of Figure~\ref{Fi:TSTcfastm}.  Notice also that that the
ratio $\delta $ between protein and mRNA decay rates is far larger in the case of 
Figure~\ref{Fi:TSTlfastm}, illustrating that, in terms of the relative stability of 
protein and mRNA, the fast mRNA approximation is much less demanding for translational
regulation than for transcriptional regulation.

\begin{figure}[h]
\includegraphics[scale=0.95]{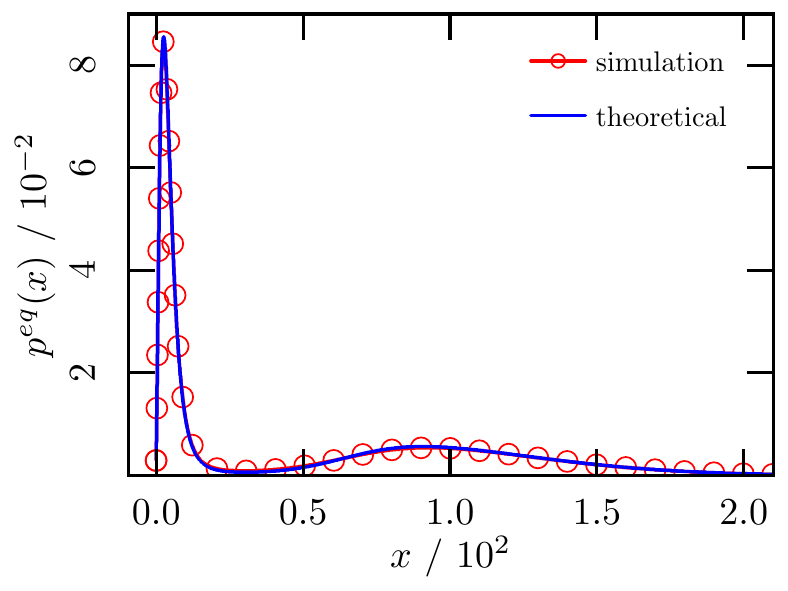}
\caption[]{(Color online) Illustration of the fast protein approximation with translational regulation.
We took $\tilde f(j,n) = f(n)$, and parameters are
$\gamma_p = 0.25$, $\mu_p = 90$,
$\gamma_m = 10$, $\mu_m = 2$, $\delta = 10^3$,
$\rho = 28$, $k = 0.1$, $a = 50$ and $\lambda = 10^{-2}$.
Error bars are smaller than markers.}
\label{Fi:TSTlfastp}
\end{figure}

For the case of translational regulation in the fast protein approximation,
the distribution of the fast species at fixed $j$ is more structured, and may be bimodal in general. Furthermore, the dependence of peak positions for the protein distribution on $j$ is also more complicated. A calculation of the parameter constraints imposed by conditions 1.\ and 2.\ in this
scenario would not lead to simple estimates such as the ones found in the previous three cases.
However, the arguments and calculations above provide the intuition that the separation regime will be reached for a certain set of parameters if $\delta$ is made large enough, as shown in Figure~\ref{Fi:TSTlfastp}. This figure also illustrates the good performance, despite the pronounced peak at low $x$, of the continuous approximation, which was used to plot the analytic curve, see Subsection  \ref{translational}. In the preceding figures the approximation for continuous $n$ is also very good (data not shown).

\section{Discussion and conclusions}  
\label{conclude}

In this paper we have established a detailed stochastic model of single-gene auto-regulation and explored its solutions when mRNA dynamics is
fast compared with protein dynamics and in the opposite regime. 
The timescale separation
allows the derivation of analytic closed form expressions for the equilibrium 
distributions of protein and mRNA. Except for  very small number of molecules, these distributions are well described in the continuous approximation, which we discuss 
in detail.
We typically find distributions that differ significantly from gaussian distributions
and exhibit heavy tails.  This is the effect of an essential ingredient of the model, the transcriptional and translational bursts, which typically have a magnitude comparable to system size.
The continuous approximation is well suited to the description of the qualitative
features of the protein equilibrium distributions as a function of the biological parameters for fast mRNA. In particular, we find that for positive auto-regulation and intermediate values of
the dimerization parameter $a$ the protein equilibrium distributions are bimodal with two non-zero peaks in a significant range of the remaining parameters.  
In more general terms, our results show that a fully stochastic description of single-gene positive
auto-regulation generates structured protein distributions that otherwise can only be explained in the framework of more complex gene regulatory networks. 

We discussed the conditions under which the timescale separations hold in good 
approximation and illustrated the performance of both regimes for transcriptional and for translational regulation by comparison with simulations. We found a broad range of parameter values where one of the two opposite timescale regimes provides a very good approximation. 
In this range, statistical measures such as mean and variance commonly used in the biological literature to characterize experimental results on protein and mRNA abundances in ensembles of cells can be readily computed from the analytic equilibrium distributions derived in Section \ref{Solanal}. However, mean and variance fall short of characterizing distributions that can be unimodal or bimodal, and non-uniformly heavy tailed. For the purpose of comparing the results of the model with real data, it is best to consider 
the full analytic distributions.

Evidence of translational regulation reported in the biological
literature raises the question of understanding its role in the context of stochastic
gene expression. Recent work has shown how different
post-transcriptional regulation mechanisms modulate
noise in protein distributions \cite{jia2010posttranscriptional}.
Here we have shown that the equilibrium protein distributions for translational 
regulation have the same form  as those that arise under transcriptional regulation in the 
case of fast mRNA. In particular, the structured protein distributions produced by
transcriptional auto-regulation with fast mRNA are also produced by translational 
auto-regulation under less demanding conditions in terms of protein and mRNA relative stability.
On the other hand, we have shown that in the translational 
regulation scenario these structured protein distributions 
are often found as equilibrium solutions also for fast protein dynamics. 
These properties suggest for translational regulation an additional biological rationale: 
it allows for efficient auto-regulation, circumventing mRNA stability.
This idea concurs with the analysis of
\cite{schwanhausser2011global}
based on experimental data for protein-mRNA lifetime pairs.

Transcriptional and translational bursts are an essential ingredient of the model
whose possible underlying mechanisms and statistics are currently being discussed in the literature. Throughout the paper, we assumed the simplest form for these bursts. Extending 
these results, in particular the validity of the timescale separation approximation, to the case of more complex mRNA and protein production statistics, see \cite{elgart2011connecting}, will be the subject
of future work.


\begin{acknowledgments}
Authors TA and AN wish to thank two anonymous referees and R. Travasso
for their useful criticism of a previous version of this paper. 
The authors are also grateful to D. Henrique for many helpful discussions
of the biological foundations and applications of the model.
Financial support from the portuguese funding agency
Funda\c c\~ao para a Ci\^encia e a Tecnologia (FCT)
under Contract POCTI/ISFL/2/261 is gratefully acknowledged. 
TA and EA were also supported by FCT under Grants PTDC-FIS-70973-2006 (TA)
and SFRH/BPD/26854/2006 (EA). 
\end{acknowledgments}


\appendix


\section{Dimer Dynamics}
\label{app:Dimers}

Consider a cell of volume $V$ where there are $n$ copies of some molecular species that can be characterized by a dimerization rate $k_d^+$ (dimensions $\text{volume.time}^{ -1 }$) and an undimerization rate $k_d^-$ (dimensions $\text{time}^{ -1 }$). Our goal here is to find the explicit form of $n_{ 2 }( n )$, the number of dimers as a function of (fixed) total copy number $n$. The equations governing dimerization dynamics of this species at fixed total density $\phi \equiv n/V$ are:
\begin{subequations}
\label{DimerDyn}
	\begin{empheq}[left=\empheqlbrace]{align} 
	\label{DimerDyn1}
		\dot{ \phi_{ 1 } } &=
			k_{ d }^{ + } \phi_{ f }^{ 2 } - k_{ d }^{ - } \phi_{ 2 } 
	\\
	\label{DimerDyn2}
		\phi &= \phi_{ f } + 2 \phi_{ 2 } \ .
	\end{empheq} 
\end{subequations}
Equation \eqref{DimerDyn1} is the rate equation for temporal dynamics, and the conservation equation \eqref{DimerDyn2} reflects that molecules are either free ($\phi_{ f } \equiv n_{ f } / V$) or bound in pairs as dimers ($\phi_{ 2 } \equiv n_{ 2 } / V$).

Defining
$k_{ d } \equiv k_{ d }^{ + } / k_{ d }^{ - }$,
equation~\eqref{DimerDyn1} yields in equilibrium:
\begin{equation}
\label{DimerEq}
	\phi_{ 2 } = k_{ d } \phi_{ f }^{ 2 } \ .
\end{equation}
Using equation~\eqref{DimerDyn2} for $\phi_{ f }$ leads, in terms of copy number, to the desired result:
\begin{equation}
\label{DimerRel}
	n_{ 2 }( n ) =
			\frac{ n }{ 2 } +
			a^2 -
			a \sqrt{
				n + a^2 } \ ,
\end{equation}
where $a$ is a dimensionless parameter defined by
$a \equiv \sqrt{ V / (8 k_d) }$.

It is also interesting to note that there are two limits in which \eqref{DimerRel} becomes very simple and intuitive. One the one hand, if $a^2 \ll n$, we find:
\begin{equation}
\label{DimerAppBound}
	n_2( n ) \approx \frac{ n }{ 2 } \ .
\end{equation}
In physical terms, this can be understood as follows: for a certain density $n/V$, if $k_{ d }$ is high enough most proteins will bind in dimers; conversely, for a certain $k_{ d }$, if density is high enough most proteins will again be bound because of increased collision probability. On the other hand, if $a^2 \gg n$, we are in the opposite limit where most proteins will be free. Taylor expansion of the square root leads in lowest order to:
\begin{equation}
\label{DimerAppFree}
	n_{ 2 }( n ) \approx
		\frac{ k_{ d } }{ V }
		n^{ 2 }  \ .
\end{equation}
This result can also be found by setting $\phi_{ f } \approx \phi$ in \eqref{DimerEq}.

\section{Mean mRNA in Equilibrium (Fast mRNA)}
\label{app:MeanmRNA}
Consider the mRNA master equation \eqref{fastmmCondMeq}. Multiplying both sides by $j$ and summing over $j$ we find an equation for the mean:
\begin{equation}
\label{mMeqMean}
\begin{aligned}
	\partial_t
		\langle id
		\rangle_{ q_{ \mid n } ( t ) } =
			&\Big[
				\beta_{ m } f( n )
				\sum_{ i \geqslant 1 }
					E_{ i }( \mu_{ m } )
					\sum_{ j \geqslant 0 }
						j ( \mathbb{ E }_m^{ -i } - 1 )
			\\
				&+ \,
				\delta_m
				\sum_{ j \geqslant 0 } j
				( \mathbb{ E }_m - 1 ) j
			\Big]
			q_{ j \mid n }( t ) \ .
\end{aligned}
\end{equation}

Let us compute (omitting the arguments $t$, $n$ for simplicity):
\begin{equation}
\label{mMeanCalc1}
\begin{aligned}
	\sum_{\substack{j \geqslant 0,\\ i \geqslant 1}}
		E_{ i }( \mu_{ m } )
		j( \mathbb{ E }_m^{ -i } - 1 )
		q_{ j } &=
	\sum_{ i \geqslant 1 }
		E_{ i }( \mu_{ m } )
		\sum_{ j \geqslant 0 } 
			j( q_{ j - i } - q_{ j } ) \; 
	\\
	&=
		\sum_{ i \geqslant 1 }
			i E_{ i }( \mu_{ m } )
			\sum_{ j \geqslant 0 }
				q_j \; 
	\\
	&=
		\mu_{ m } \ ,
\end{aligned}
\end{equation}
where we have made use of the fact that $q_j = 0$ whenever copy number $j$ is negative.
Now let us look at:
\begin{equation}
\label{mMeanCalc2}
\begin{aligned}
	\sum_{ j \geqslant 0 } j
		( \mathbb{ E }_m - 1 )
		j q_{ j } \;
	&=
		\sum_{ j \geqslant 0 }
			j ( j + 1 ) q_{ j + 1 } -
		\sum_{ j \geqslant 0 }
			j^2 q_j \; 
	\\
	&=
		\sum_{ j \geqslant 1 }
			( j - 1 ) j q_j -
		\sum_{ j \geqslant 0 }
			j^2 q_j \; 
	\\
	&=
		- \sum_{ j \geqslant 0 }
			j q_j \; 
	=
		- \langle id \rangle_{ q_{ \mid n } ( t ) } \ .
\end{aligned}
\end{equation}

Since we are looking for the equilibrium mean we now set the left-hand side of~\eqref{mMeqMean} to zero, and using results~\eqref{mMeanCalc1} and~\eqref{mMeanCalc2} we find the desired result:
\begin{equation}
\label{mMean}
	\langle id \rangle_{ q^{ eq }_{ \mid n } } =
		\mu_{ m } \gamma_{ m } f( n ) \ .
\end{equation}


\section{Continuous Approximation}
\label{app:ContApprox}

Here we study a continuous approximation for equations of the form:
\begin{equation}
\label{toContinuous}
	\dot{ p }_n( t ) =
		\Big[
			r \delta 
			\sum_{ i \geqslant 1 }
			E_{ i }( \mu )
				( \mathbb{ E }^{ -i } - 1 ) f( n ) +
			\delta( \mathbb{ E } - 1 ) n
		\Big]
		p_n( t ) \ ,
\end{equation}
where $f$ is some function of (protein or mRNA) copy number $n$, $r \neq 0$ and $\delta \neq 0$ are constants, and the step operator $\mathbb{ E }$ raises $n$. For some time $t$, let copy number $n$ be fixed, and let $x = \lambda n$ be the corresponding concentration. In accordance with the main text, the convention $f( n ) = f( x )$ will be used. First, note that a reasonable definition for the continuous distribution obeys:
\begin{equation}
\label{pContDef}
\begin{aligned}
p_n( t ) &\equiv
		p( x , t ) \lambda \Big[ ( n + 1/2 ) - ( n - 1/2 ) \Big]
		\\
		&
		\approx
		\int_{ n - 1/2 }^{ n + 1/2 }
		p(x,t)\ dx = \lambda p( x , t ) \ .
\end{aligned}
\end{equation}

Now consider the conditioned geometric distribution. We have:
\begin{equation}
\label{approxGeometric1}
	E_n( \mu ) =
		\frac { ( \mu - 1 )^{( n - 1 ) }}
			{ \mu^n } \; = \frac{ 1 }{ \mu - 1 }
		e^{ - n \log ( 1 - 1 / \mu ) } \ .
\end{equation}
If we take $\mu \gg 1$ (which is biologically common, especially for proteins, see for example \cite{kaufmann2007stochastic,taniguchi2010quantifying}) and expand $\log(1 - 1 / \mu)$ around $1/\mu = 0$ we find to lowest order:
\begin{equation}
\label{approxGeometric2}
	E_n( \mu )  \approx
		\frac { 1 }{ \mu }
			e^{ - n / \mu} =
		\lambda \frac { 1 }{ \tilde\mu }
			e^{ - x / \tilde\mu } =
		\lambda E( x , \tilde\mu )\ ,
\end{equation}
with $\tilde\mu = \lambda \mu$.
Now notice that, apart from constant coefficients, the creation term in equation~\eqref{toContinuous} may be written:
\begin{equation}
\label{Creation}
\begin{aligned}
	& \sum_{ i \geqslant 1 }
		E_{ i }( \mu )
		( \mathbb{ E }^{ -i } - 1 ) f( n ) p_n( t )  
		\\
	 & = \sum_{ i \geqslant 1 }
		E_{ i }( \mu )
		f( n - i ) p_{ n - i } -f( n ) p_n( t )
	\\		
	& = \sum_{ i = 0 }^{ n }
		\left(
			E_{ n - i }( \mu ) - \delta_{n , i }
		\right)
		f( i ) p_i \ ,
\end{aligned}
\end{equation}
where $\delta_{n , i }$ is a Kronecker Delta symbol. Note that the upper limit of the sum can be extended to infinity by taking $E_j( \mu )$ = 0 for $j \leqslant 0$, and the lower limit can be extended to negative infinity since $p_i = 0$ for $i < 0$.

The Kronecker Delta term reads:
\begin{equation}
\label{approxKronecker}
\begin{aligned}
	\sum_{ i = 0 }^{ n }
		 \delta_{n , i } f( i ) p_i &= f( n ) p_n( t ) 
	\\
	& = \lambda \int_0^{ x } \delta_D( x - y) f( y ) p( y, t ) \, dy \ ,
\end{aligned}
\end{equation}
where $\delta_D$ is the Dirac Delta. Notice that, for a meaningful conversion to the continuous case, the lower limit of the integral must be strictly included (in order to encompass the contribution of the Delta function). Thus, the upper and lower limits of the integral may be extended to infinity.

For the conditioned geometric distribution term in \eqref{Creation} we may write:
\begin{equation}
\label{approxGeometric3}
\begin{aligned}
	\sum_{ i = 0 }^{ n }
		E_{ n - i }( \mu )	
		f( i ) p_i &\approx 
	\sum_{ i = 0 }^{ n }
		\lambda E( \lambda ( n - i ) , \tilde\mu )	
		f( i ) \lambda p( \lambda i, t )  
	\\
	&\approx
	\lambda \int_0^{ x }
		E( x - y, \tilde\mu ) f( y ) p( y , t ) \, dy \ ,
\end{aligned}
\end{equation}
where again the upper and lower limits of the integral may be extended do plus and minus infinity by considering, respectively, $E(y, \tilde\mu) = 0$ and $p( y , t ) = 0$ for negative $y$. Here, the approximations $\mu \gg 1$ (approximating the conditioned geometric distribution with an exponential distribution) and $\lambda \ll 1$ (approximating the sum with an integral, i.e.~considering $x$ continuous) have been explicitly used.

Finally, the degradation term in equation~\eqref{toContinuous} reads, apart from a factor of~$\delta$:
\begin{equation}
\begin{aligned}
	( \mathbb{ E } - 1 ) n p_n( t ) &=
	\Big[
		( n + 1 ) p_{ n + 1 }( t ) -  n p_n( t )
	\Big] 
	\\
	&=
	\frac{ 1 }{ \lambda }
	\Big[
		( x + \lambda ) \lambda p( x + \lambda )( t ) -  x \lambda p( x ,  t )
	\Big] 
	\\
	&\approx
	\lambda \partial_x ( x p( x , t ) ) \ ,
\end{aligned}
\end{equation}
where we again make use of $\lambda \ll 1$ to approximate a finite difference with a derivative. Noting that $\dot{ p }_n ( t ) = \lambda \dot{ p } ( x , t )$ and collecting terms we find:
\begin{equation}
\label{Cont}
\begin{aligned}
	\dot{ p }( x , t ) =
		& \, r \delta
		\int_{ 0 }^{ x }
			f( y )
			\Big[
				E( x - y  , \tilde\mu ) - \delta_D ( x - y )
			\Big] p( y, t) \, dy 
		\\
		&+ \delta \, \partial_{ x }
			\Big[
				x p( x , t )
			\Big] \ .
\end{aligned}
\end{equation}


\section{Continuous Equilibrium Distributions}
\label{app:ContSol}

Here we follow~\cite{friedman2006linking} to obtain an analytical solution to equation~\eqref{Cont}. As discussed in Appendix~\ref{app:ContApprox}, the upper and lower integration limits may be extended to plus and minus infinity, respectively. Thus, defining:
\begin{equation}
	w( x \, , \tilde\mu ) = E( x , \tilde\mu ) - \delta_D( x ) \; ,
\end{equation}
we may write:
\begin{equation}
	\dot{ p }( x , t ) =
		r \delta 
		( w( \tilde\mu ) * f p ) ( x , t ) +
		\delta \, \partial_x
			\Big[
				x p( x , t )
			\Big] \ ,
\end{equation}
where $*$ is a convolution product. In equilibrium we have:
\begin{equation}
	- \partial_{ x }
	\Big[
		x p^{ eq }( x )
	\Big] =
	r
	( w( \tilde\mu ) * f p^{ eq } ) ( x )\ .
\end{equation}
Laplace transformation of this equation leads to:
\begin{equation}
\begin{aligned}
	s \ \partial_s \hat p( s ) &= r \, \hat w( s ) \, \mathcal{L} ( f p^{ eq } )( s ) 
	\\
	&= r \, \hat w( s ) ( \hat f * \hat p )( s )
	\\
	&= - r \frac{ s }{ s + 1 / \tilde\mu } ( \hat f * \hat p )( s ) \ .
\end{aligned}
\end{equation}
Here,
$\hat g ( s )= \mathcal{L}( g )( s ) = \int_{ 0 }^{ +\infty } e^{ -s x } g( x ) \, dx$ (integration limit $0$ strictly included)
is the Laplace transform of function $g$ (evaluated at $s$), and $\hat p = \mathcal{L}( p^{ eq } )$. Convolution theorems have been used in the first and second lines, and in the third line the explicit form of $\hat w ( s )$ was substituted. Rearranging terms we have:
\begin{equation}
	( s + 1 / \tilde\mu ) \hat p( s ) =  - r ( \hat f * \hat p )( s ) \ ,
\end{equation}
which inverse-transforms to:
\begin{equation}
\label{tosolve}
	\partial_x [ x p^{ eq }( x ) ] =
	 \left(
	 	r f( x ) / x - 1 / \tilde\mu
	\right) x p^{ eq }( x ) \; .
\end{equation}
This equation can easily be solved, leading to:
\begin{equation}
	p^{ eq }( x ) =
		A_c \, x^{ -1 } e^{ - x / \tilde\mu }
		e^{ r \int_c^{ x } du f( u ) / u } \  .
\end{equation}
The constant $A_c$ is determined by normalization (depending on the arbitrary integration limit $c$).

Consider now the case $f( x ) = 1$, for all $x$.  Solving the integral in \eqref{fastmpContSol} and normalizing the probability distribution to integral unity we find:
\begin{equation}
\begin{aligned}
	p^{ eq }( x ) &=
	\frac{ x^{ r -1 } e^{ - x / \tilde\mu } }
		{ \tilde\mu^{ r } \Gamma( r ) }
	\\
	&= G( x , r, \tilde\mu) \ .
\end{aligned}
\end{equation}
This is the Gamma distribution of parameters $r$ and $\tilde\mu$ ($\Gamma$ is the Euler Gamma function). With $r = \mu_{ m } \gamma_{ m } \gamma_p$ and $\tilde\mu$ the mean rescaled protein burst size (with definitions according to the main text), this is the equilibrium solution for unregulated protein dynamics with fast mRNA.

\section{Discrete Equilibrium Distributions}
\label{app:Sol}

In this Appendix we analyze, directly in the discrete setting, equation~\eqref{toContinuous}. Analogously to the continuous case, the discrete Master Equation may be written:
\begin{equation}
	\dot{ p }_n( t ) =
	r \delta 
	( w( \mu ) * f p ) ( n ) +
	\delta
	\Big[
		( n + 1 ) p_{ n + 1 }( t ) -  n p_n( t )
	\Big] \ ,
\end{equation}
where $*$ is now the discrete convolution product, and $w( n \, , \mu ) = E_{ n }( \mu ) - \delta_{ n , 0 } $.
We now follow the procedures of Appendix~\ref{app:ContSol} using the Z transform instead of the Laplace transform,
$\hat g( s )~=~\mathcal{ Z }( g ) ( s )= \sum_{ n = 0 }^{ +\infty } s^{ -n } g( n )$, $\mathcal{ Z }( p^{ eq } ) = \hat p$. The corresponding equation in ``momentum space" is:
\begin{equation}
	s ( s - 1 ) \partial_s \hat p( s ) + \frac{ s }{ \mu } \partial_s \hat p( s ) =
	- r \left(
		\hat f * \hat p
		\right) ( s ) \ .
\end{equation}
Inverse-transforming, we get:
\begin{equation}
	( n + 1 ) p_{ n + 1 }^{ eq } + ( 1 / \mu - 1 ) n p_n^{ eq } = r f( n ) p_n^{ eq } \ ,
\end{equation}
leading to the recurrence relation:
\begin{equation}
\begin{cases}
	& p_1^{ eq } = r f( 0 ) p_0^{ eq }, \\
	& ( n + 1 ) p_{ n + 1 } =
	\left(
		r \frac{ f( n ) }{ n } + \frac{ \mu - 1 }{ \mu }
	\right)
	n p_n^{ eq }, \ n \geqslant 1 \ .
\end{cases}
\end{equation}
This is easily solved, yielding:
\begin{equation}
\label{sol}
	p_n^{ eq } =
	\frac{ r f( 0 ) p_0^{ eq } }{ n }
	\prod_{ i = 1 }^{ n - 1 }
	\left(
			r \frac{ f( i ) }{ i } + \frac{ \mu - 1 }{ \mu }
	\right) \ ,
\end{equation}
for all $n \geqslant 1$, with $p_0^{ eq }$ determined by normalization (and the standard convention that the product equals one when the upper limit is smaller than the lower). Note that if $f$ is a regulation function as per the main text we have $f( 0 ) = 1$, since the promoter is necessarily free when no protein is present.

Consider now the case $f( n ) = 1$ for all $n$. Write \eqref{sol} as:
\begin{equation}
\begin{aligned}
	p_n^{ eq } &=
	\frac{ \mu }{ \mu - 1 } \frac{ r f( 0 ) p_0^{ eq } }{ n }
	\left(
		\frac{ \mu - 1 }{ \mu }
	\right)^n \; 
	\prod_{ i = 1 }^{ n - 1 }
	\left(
		\frac{ \mu }{ \mu - 1 } r \frac{ f( i ) }{ i } + 1
	\right)
	\\
	&=
	\frac{ r' f( 0 ) p_0^{ eq } }{ n }
	\left(
		\frac{ \mu - 1 }{ \mu }
	\right)^n  \; 
	\prod_{ i = 1 }^{ n - 1 }
	\left(
		r' \frac{ f( i ) }{ i } + 1
	\right) \ ,
\end{aligned}
\end{equation}
with $r' = r  \mu / ( \mu - 1 )$. The product can be solved explicitly is terms of Gamma functions, and normalizing to unit sum we find:
\begin{equation}
\begin{aligned}
	p_n^{ eq } &= \frac{ 1 }{ \mu^{r'} }
	\left(
		\frac{ \mu - 1 }{ \mu }
	\right)^n
	\frac{ \Gamma( n + r' ) }{ \Gamma( r' ) \Gamma( n + 1 ) }
	\\
	&= N_n \left( r', \frac{ 1 }{ \mu }
		\right) \ .
\end{aligned}
\end{equation}
This is the Negative Binomial distribution of parameters $\gamma'$ and $1/\mu$. The parameters are defined such that:
\begin{equation}
	N_n( k, p ) = p^k (1 - p)^n \binom{ n + k - 1 }{ k - 1 } \ .
\end{equation}
As in the continuous case (Appendix~\ref{app:ContSol}), with $r = \mu_{ m } \gamma_{ m } \gamma_{ p}$ and $\mu$ the mean protein burst size $\mu_p$ (definitions according to the main text), this is the discrete solution for unregulated protein dynamics with fast mRNA (as reported for example in~\cite{shahrezaei2008analytical}).\\\\


\section{Bimodal Equilibrium Protein Distributions}
\label{app:Quartic}

Consider the continuous equilibrium distribution for protein with fast mRNA, given by \eqref{fastmpContSol}. The derivative of this probability distribution is given by:
\begin{equation}
\label{der}
	\partial_x p^{eq}(x) =
	\Big[
		r \tilde\mu f(x) - (x + \tilde\mu)
	\Big]
	\frac{ p^{eq}(x) }{ \tilde\mu x } \ .
\end{equation}
If $p^{eq}$ peaks at zero (i.e.\ if $\partial_x p^{eq}(0) < 0$), the term in brackets in equation~\eqref{der} must be negative at zero. Because $p^{eq}(x) > 0$ for all $x > 0$, other extrema of $p^{eq}$ must satisfy:
\begin{equation}
\label{ref}
	r \tilde\mu f(x) - (x + \tilde\mu) = 0 \ .
\end{equation}
Consider $f(x)$ as given by~\eqref{fApprox}. A change of variables to $z = \sqrt{x + \tilde a^2}$ in equation~\eqref{ref} leads to an equivalent quartic equation, 
\begin{equation}
	P_4(z) = - z^4 + 2 \tilde a z^3 + \alpha_2 z^2 + \alpha_1 z + \alpha_0 \; =\ 0\ ,
\end{equation}
where the $\alpha_i$ are real constants determined by the biological parameters.
The equation $P''_4(z)=0$ is quadratic in $z$ and has the two solutions:
\begin{equation}
	z = \frac{ \tilde a }{ 2 } \pm \sqrt{\left(\frac{ \tilde a }{ 2 }\right)^2 + \frac{ \alpha_2 }{ 6 }} \ .
\end{equation}
If they are real, one of these solutions necessarily obeys $z < \tilde a$. Therefore, 
$P''_4(z)$ has at most one root in $z > \tilde a$.

We now proceed to prove that $p^{eq}$ is at most bimodal. Since zeros of $P_4$ correspond alternately to 
maxima and minima of $p^{eq}$, the presence of more than two maxima requires at least four positive 
roots of $P_4(z)$ in $z > \tilde a$. But then  $P''_4(z)$ would have at least two roots in 
$z > \tilde a$.


\begin{thebibliography}{99}

\expandafter\ifx\csname natexlab\endcsname\relax\def\natexlab#1{#1}\fi
\expandafter\ifx\csname bibnamefont\endcsname\relax
  \def\bibnamefont#1{#1}\fi
\expandafter\ifx\csname bibfnamefont\endcsname\relax
  \def\bibfnamefont#1{#1}\fi
\expandafter\ifx\csname citenamefont\endcsname\relax
  \def\citenamefont#1{#1}\fi
\expandafter\ifx\csname url\endcsname\relax
  \def\url#1{\texttt{#1}}\fi
\expandafter\ifx\csname urlprefix\endcsname\relax\def\urlprefix{URL }\fi
\providecommand{\bibinfo}[2]{#2}
\providecommand{\eprint}[2][]{\url{#2}}

\bibitem[{\citenamefont{Berg and Purcell}(1977)}]{berg1977physics}
\bibinfo{author}{\bibfnamefont{H.}~\bibnamefont{Berg}} \bibnamefont{and}
  \bibinfo{author}{\bibfnamefont{E.}~\bibnamefont{Purcell}},
  \bibinfo{journal}{Biophysical Journal} \textbf{\bibinfo{volume}{20}},
  \bibinfo{pages}{193} (\bibinfo{year}{1977}), ISSN \bibinfo{issn}{0006-3495}.

\bibitem[{\citenamefont{Berg}(1978)}]{berg1978model}
\bibinfo{author}{\bibfnamefont{O.}~\bibnamefont{Berg}},
  \bibinfo{journal}{Journal of Theoretical Biology}
  \textbf{\bibinfo{volume}{71}}, \bibinfo{pages}{587} (\bibinfo{year}{1978}).

\bibitem[{\citenamefont{Bialek and Setayeshgar}(2005)}]{bialek2005physical}
\bibinfo{author}{\bibfnamefont{W.}~\bibnamefont{Bialek}} \bibnamefont{and}
  \bibinfo{author}{\bibfnamefont{S.}~\bibnamefont{Setayeshgar}},
  \bibinfo{journal}{Proceedings of the National Academy of Sciences of the
  United States of America} \textbf{\bibinfo{volume}{102}},
  \bibinfo{pages}{10040} (\bibinfo{year}{2005}).

\bibitem[{\citenamefont{Bialek and
  Setayeshgar}(2008)}]{bialek2008cooperativity}
\bibinfo{author}{\bibfnamefont{W.}~\bibnamefont{Bialek}} \bibnamefont{and}
  \bibinfo{author}{\bibfnamefont{S.}~\bibnamefont{Setayeshgar}},
  \bibinfo{journal}{Physical Review Letters} \textbf{\bibinfo{volume}{100}},
  \bibinfo{pages}{258101} (\bibinfo{year}{2008}).

\bibitem[{\citenamefont{Fraser et~al.}(2004)\citenamefont{Fraser, Hirsh,
  Giaever, Kumm, and Eisen}}]{fraser2004noise}
\bibinfo{author}{\bibfnamefont{H.}~\bibnamefont{Fraser}},
  \bibinfo{author}{\bibfnamefont{A.}~\bibnamefont{Hirsh}},
  \bibinfo{author}{\bibfnamefont{G.}~\bibnamefont{Giaever}},
  \bibinfo{author}{\bibfnamefont{J.}~\bibnamefont{Kumm}}, \bibnamefont{and}
  \bibinfo{author}{\bibfnamefont{M.}~\bibnamefont{Eisen}},
  \bibinfo{journal}{PLoS Biology} \textbf{\bibinfo{volume}{2}},
  \bibinfo{pages}{e137} (\bibinfo{year}{2004}), ISSN \bibinfo{issn}{1545-7885}.

\bibitem[{\citenamefont{Elowitz et~al.}(2002)\citenamefont{Elowitz, Levine,
  Siggia, and Swain}}]{elowitz2002stochastic}
\bibinfo{author}{\bibfnamefont{M.}~\bibnamefont{Elowitz}},
  \bibinfo{author}{\bibfnamefont{A.}~\bibnamefont{Levine}},
  \bibinfo{author}{\bibfnamefont{E.}~\bibnamefont{Siggia}}, \bibnamefont{and}
  \bibinfo{author}{\bibfnamefont{P.}~\bibnamefont{Swain}},
  \bibinfo{journal}{Science} \textbf{\bibinfo{volume}{297}},
  \bibinfo{pages}{1183} (\bibinfo{year}{2002}).

\bibitem[{\citenamefont{Friedman et~al.}(2006)\citenamefont{Friedman, Cai, and
  Xie}}]{friedman2006linking}
\bibinfo{author}{\bibfnamefont{N.}~\bibnamefont{Friedman}},
  \bibinfo{author}{\bibfnamefont{L.}~\bibnamefont{Cai}}, \bibnamefont{and}
  \bibinfo{author}{\bibfnamefont{X.}~\bibnamefont{Xie}},
  \bibinfo{journal}{Physical Review Letters} \textbf{\bibinfo{volume}{97}},
  \bibinfo{pages}{168302} (\bibinfo{year}{2006}), ISSN
  \bibinfo{issn}{1079-7114}.

\bibitem[{\citenamefont{Kalmar et~al.}(2009)\citenamefont{Kalmar, Lim, Hayward,
  Mu{\~n}oz-Descalzo, Nichols, Garcia-Ojalvo, and
  Martinez~Arias}}]{kalmar2009regulated}
\bibinfo{author}{\bibfnamefont{T.}~\bibnamefont{Kalmar}},
  \bibinfo{author}{\bibfnamefont{C.}~\bibnamefont{Lim}},
  \bibinfo{author}{\bibfnamefont{P.}~\bibnamefont{Hayward}},
  \bibinfo{author}{\bibfnamefont{S.}~\bibnamefont{Mu{\~n}oz-Descalzo}},
  \bibinfo{author}{\bibfnamefont{J.}~\bibnamefont{Nichols}},
  \bibinfo{author}{\bibfnamefont{J.}~\bibnamefont{Garcia-Ojalvo}},
  \bibnamefont{and}
  \bibinfo{author}{\bibfnamefont{A.}~\bibnamefont{Martinez~Arias}},
  \bibinfo{journal}{PLoS Biology} \textbf{\bibinfo{volume}{7}},
  \bibinfo{pages}{e1000149} (\bibinfo{year}{2009}), ISSN
  \bibinfo{issn}{1545-7885}.

\bibitem[{\citenamefont{Ru{\'e} and Garcia-Ojalvo}(2011)}]{rue2011gene}
\bibinfo{author}{\bibfnamefont{P.}~\bibnamefont{Ru{\'e}}} \bibnamefont{and}
  \bibinfo{author}{\bibfnamefont{J.}~\bibnamefont{Garcia-Ojalvo}},
  \bibinfo{journal}{Mathematical Biosciences} \textbf{\bibinfo{volume}{231}},
  \bibinfo{pages}{90} (\bibinfo{year}{2011}).

\bibitem[{\citenamefont{Cai et~al.}(2006)\citenamefont{Cai, Friedman, and
  Xie}}]{cai2006stochastic}
\bibinfo{author}{\bibfnamefont{L.}~\bibnamefont{Cai}},
  \bibinfo{author}{\bibfnamefont{N.}~\bibnamefont{Friedman}}, \bibnamefont{and}
  \bibinfo{author}{\bibfnamefont{X.}~\bibnamefont{Xie}},
  \bibinfo{journal}{Nature} \textbf{\bibinfo{volume}{440}},
  \bibinfo{pages}{358} (\bibinfo{year}{2006}).

\bibitem[{\citenamefont{Kaufmann and van
  Oudenaarden}(2007)}]{kaufmann2007stochastic}
\bibinfo{author}{\bibfnamefont{B.}~\bibnamefont{Kaufmann}} \bibnamefont{and}
  \bibinfo{author}{\bibfnamefont{A.}~\bibnamefont{van Oudenaarden}},
  \bibinfo{journal}{Current Opinion in Genetics \& Development}
  \textbf{\bibinfo{volume}{17}}, \bibinfo{pages}{107} (\bibinfo{year}{2007}).

\bibitem[{\citenamefont{Ozbudak et~al.}(2002)\citenamefont{Ozbudak, Thattai,
  Kurtser, Grossman, and van Oudenaarden}}]{ozbudak2002regulation}
\bibinfo{author}{\bibfnamefont{E.}~\bibnamefont{Ozbudak}},
  \bibinfo{author}{\bibfnamefont{M.}~\bibnamefont{Thattai}},
  \bibinfo{author}{\bibfnamefont{I.}~\bibnamefont{Kurtser}},
  \bibinfo{author}{\bibfnamefont{A.}~\bibnamefont{Grossman}}, \bibnamefont{and}
  \bibinfo{author}{\bibfnamefont{A.}~\bibnamefont{van Oudenaarden}},
  \bibinfo{journal}{Nature Genetics} \textbf{\bibinfo{volume}{31}},
  \bibinfo{pages}{69} (\bibinfo{year}{2002}).

\bibitem[{\citenamefont{Taniguchi et~al.}(2010)\citenamefont{Taniguchi, Choi,
  Li, Chen, Babu, Hearn, Emili, and Xie}}]{taniguchi2010quantifying}
\bibinfo{author}{\bibfnamefont{Y.}~\bibnamefont{Taniguchi}},
  \bibinfo{author}{\bibfnamefont{P.}~\bibnamefont{Choi}},
  \bibinfo{author}{\bibfnamefont{G.}~\bibnamefont{Li}},
  \bibinfo{author}{\bibfnamefont{H.}~\bibnamefont{Chen}},
  \bibinfo{author}{\bibfnamefont{M.}~\bibnamefont{Babu}},
  \bibinfo{author}{\bibfnamefont{J.}~\bibnamefont{Hearn}},
  \bibinfo{author}{\bibfnamefont{A.}~\bibnamefont{Emili}}, \bibnamefont{and}
  \bibinfo{author}{\bibfnamefont{X.}~\bibnamefont{Xie}},
  \bibinfo{journal}{Science} \textbf{\bibinfo{volume}{329}},
  \bibinfo{pages}{533} (\bibinfo{year}{2010}).

\bibitem[{\citenamefont{Suter et~al.}(2011)\citenamefont{Suter, Molina,
  Gatfield, Schneider, Schibler, and Naef}}]{suter2011mammalian}
\bibinfo{author}{\bibfnamefont{D.}~\bibnamefont{Suter}},
  \bibinfo{author}{\bibfnamefont{N.}~\bibnamefont{Molina}},
  \bibinfo{author}{\bibfnamefont{D.}~\bibnamefont{Gatfield}},
  \bibinfo{author}{\bibfnamefont{K.}~\bibnamefont{Schneider}},
  \bibinfo{author}{\bibfnamefont{U.}~\bibnamefont{Schibler}}, \bibnamefont{and}
  \bibinfo{author}{\bibfnamefont{F.}~\bibnamefont{Naef}},
  \bibinfo{journal}{Science} \textbf{\bibinfo{volume}{322}},
  \bibinfo{pages}{472} (\bibinfo{year}{2011}).

\bibitem[{\citenamefont{Larson et~al.}(2011)\citenamefont{Larson, Zenklusen,
  Wu, Chao, and Singer}}]{larson2011realtime}
\bibinfo{author}{\bibfnamefont{D.}~\bibnamefont{Larson}},
  \bibinfo{author}{\bibfnamefont{D.}~\bibnamefont{Zenklusen}},
  \bibinfo{author}{\bibfnamefont{B.}~\bibnamefont{Wu}},
  \bibinfo{author}{\bibfnamefont{J.}~\bibnamefont{Chao}}, \bibnamefont{and}
  \bibinfo{author}{\bibfnamefont{R.}~\bibnamefont{Singer}},
  \bibinfo{journal}{Science} \textbf{\bibinfo{volume}{322}},
  \bibinfo{pages}{475} (\bibinfo{year}{2011}).

\bibitem[{\citenamefont{Hornos et~al.}(2009)\citenamefont{Hornos, Schultz,
  Innocentini, Wang, , Walczak, Onuchic, and
  Wolynes}}]{hornos2005selfregulating}
\bibinfo{author}{\bibfnamefont{J.}~\bibnamefont{Hornos}},
  \bibinfo{author}{\bibfnamefont{D.}~\bibnamefont{Schultz}},
  \bibinfo{author}{\bibfnamefont{G.}~\bibnamefont{Innocentini}},
  \bibinfo{author}{\bibfnamefont{J.}~\bibnamefont{Wang}}, ,
  \bibinfo{author}{\bibfnamefont{A.}~\bibnamefont{Walczak}},
  \bibinfo{author}{\bibfnamefont{J.}~\bibnamefont{Onuchic}}, \bibnamefont{and}
  \bibinfo{author}{\bibfnamefont{P.}~\bibnamefont{Wolynes}},
  \bibinfo{journal}{Physical Review E} \textbf{\bibinfo{volume}{72}},
  \bibinfo{pages}{051907} (\bibinfo{year}{2009}).

\bibitem[{\citenamefont{Mugler et~al.}(2009)\citenamefont{Mugler, Walczak, and
  Wiggins}}]{mugler2009spectral}
\bibinfo{author}{\bibfnamefont{A.}~\bibnamefont{Mugler}},
  \bibinfo{author}{\bibfnamefont{A.}~\bibnamefont{Walczak}}, \bibnamefont{and}
  \bibinfo{author}{\bibfnamefont{C.}~\bibnamefont{Wiggins}},
  \bibinfo{journal}{Physical Review E} \textbf{\bibinfo{volume}{80}},
  \bibinfo{pages}{041921} (\bibinfo{year}{2009}).

\bibitem[{\citenamefont{Iyer-Biswas et~al.}(2009)\citenamefont{Iyer-Biswas,
  Hayot, and Jayaprakash}}]{iyer-biswas2009stochasticity}
\bibinfo{author}{\bibfnamefont{S.}~\bibnamefont{Iyer-Biswas}},
  \bibinfo{author}{\bibfnamefont{F.}~\bibnamefont{Hayot}}, \bibnamefont{and}
  \bibinfo{author}{\bibfnamefont{C.}~\bibnamefont{Jayaprakash}},
  \bibinfo{journal}{Physical Review E} \textbf{\bibinfo{volume}{79}},
  \bibinfo{pages}{031911} (\bibinfo{year}{2009}).

\bibitem[{\citenamefont{Assaf et~al.}(2011)\citenamefont{Assaf, Roberts, and
  Luthey-Schulten}}]{assaf2011determining}
\bibinfo{author}{\bibfnamefont{M.}~\bibnamefont{Assaf}},
  \bibinfo{author}{\bibfnamefont{E.}~\bibnamefont{Roberts}}, \bibnamefont{and}
  \bibinfo{author}{\bibfnamefont{Z.}~\bibnamefont{Luthey-Schulten}},
  \bibinfo{journal}{Physical Review Letters} \textbf{\bibinfo{volume}{106}},
  \bibinfo{pages}{248102} (\bibinfo{year}{2011}).

\bibitem[{\citenamefont{Waters and Storz}(2009)}]{waters2009regulatory}
\bibinfo{author}{\bibfnamefont{L.}~\bibnamefont{Waters}} \bibnamefont{and}
  \bibinfo{author}{\bibfnamefont{G.}~\bibnamefont{Storz}},
  \bibinfo{journal}{Cell} \textbf{\bibinfo{volume}{136}}, \bibinfo{pages}{615}
  (\bibinfo{year}{2009}).

\bibitem[{\citenamefont{Wang et~al.}(2008)\citenamefont{Wang, Levasseur, and
  Orkin}}]{wang2008requirement}
\bibinfo{author}{\bibfnamefont{J.}~\bibnamefont{Wang}},
  \bibinfo{author}{\bibfnamefont{D.}~\bibnamefont{Levasseur}},
  \bibnamefont{and} \bibinfo{author}{\bibfnamefont{S.}~\bibnamefont{Orkin}},
  \bibinfo{journal}{Proceedings of the National Academy of Sciences of the
  United States of America} \textbf{\bibinfo{volume}{105}},
  \bibinfo{pages}{6326} (\bibinfo{year}{2008}).

\bibitem[{\citenamefont{Schwanh{\"a}usser
  et~al.}(2011)\citenamefont{Schwanh{\"a}usser, Busse, Li, Dittmar,
  Schuchhardt, Wolf, Chen, and Selbach}}]{schwanhausser2011global}
\bibinfo{author}{\bibfnamefont{B.}~\bibnamefont{Schwanh{\"a}usser}},
  \bibinfo{author}{\bibfnamefont{D.}~\bibnamefont{Busse}},
  \bibinfo{author}{\bibfnamefont{N.}~\bibnamefont{Li}},
  \bibinfo{author}{\bibfnamefont{G.}~\bibnamefont{Dittmar}},
  \bibinfo{author}{\bibfnamefont{J.}~\bibnamefont{Schuchhardt}},
  \bibinfo{author}{\bibfnamefont{J.}~\bibnamefont{Wolf}},
  \bibinfo{author}{\bibfnamefont{W.}~\bibnamefont{Chen}}, \bibnamefont{and}
  \bibinfo{author}{\bibfnamefont{M.}~\bibnamefont{Selbach}},
  \bibinfo{journal}{Nature} \textbf{\bibinfo{volume}{473}},
  \bibinfo{pages}{337} (\bibinfo{year}{2011}).

\bibitem[{\citenamefont{Guantes and Poyatos}(2008)}]{guantes2008multistable}
\bibinfo{author}{\bibfnamefont{R.}~\bibnamefont{Guantes}} \bibnamefont{and}
  \bibinfo{author}{\bibfnamefont{J.}~\bibnamefont{Poyatos}},
  \bibinfo{journal}{PLoS Computational Biology} \textbf{\bibinfo{volume}{4}},
  \bibinfo{pages}{e1000235} (\bibinfo{year}{2008}).

\bibitem[{\citenamefont{van Kampen}(1992)}]{van1992stochastic}
\bibinfo{author}{\bibfnamefont{N.}~\bibnamefont{van Kampen}},
  \emph{\bibinfo{title}{Stochastic processes in physics and chemistry}}
  (\bibinfo{publisher}{North Holland}, \bibinfo{year}{1992}).

\bibitem[{\citenamefont{Liu et~al.}(1993)\citenamefont{Liu, Li, and
  Giddings}}]{liu1993rapid}
\bibinfo{author}{\bibfnamefont{M.}~\bibnamefont{Liu}},
  \bibinfo{author}{\bibfnamefont{P.}~\bibnamefont{Li}}, \bibnamefont{and}
  \bibinfo{author}{\bibfnamefont{J.}~\bibnamefont{Giddings}},
  \bibinfo{journal}{Protein Science: A Publication of the Protein Society}
  \textbf{\bibinfo{volume}{2}}, \bibinfo{pages}{1520} (\bibinfo{year}{1993}).

\bibitem[{\citenamefont{Xu et~al.}(1998)\citenamefont{Xu, Tsai, and
  Nussinov}}]{xu1998mechanism}
\bibinfo{author}{\bibfnamefont{D.}~\bibnamefont{Xu}},
  \bibinfo{author}{\bibfnamefont{C.}~\bibnamefont{Tsai}}, \bibnamefont{and}
  \bibinfo{author}{\bibfnamefont{R.}~\bibnamefont{Nussinov}},
  \bibinfo{journal}{Protein Science: A Publication of the Protein Society}
  \textbf{\bibinfo{volume}{7}}, \bibinfo{pages}{533} (\bibinfo{year}{1998}).

\bibitem[{\citenamefont{B\'enichou et~al.}(2009)\citenamefont{B\'enichou,
  Kafri, Sheinman, and Voituriez}}]{benichou2009searching}
\bibinfo{author}{\bibfnamefont{O.}~\bibnamefont{B\'enichou}},
  \bibinfo{author}{\bibfnamefont{Y.}~\bibnamefont{Kafri}},
  \bibinfo{author}{\bibfnamefont{M.}~\bibnamefont{Sheinman}}, \bibnamefont{and}
  \bibinfo{author}{\bibfnamefont{R.}~\bibnamefont{Voituriez}},
  \bibinfo{journal}{Physical Review Letters} \textbf{\bibinfo{volume}{103}},
  \bibinfo{pages}{138102} (\bibinfo{year}{2009}).

\bibitem[{\citenamefont{Paulsson}(2005)}]{paulsson2005models}
\bibinfo{author}{\bibfnamefont{J.}~\bibnamefont{Paulsson}},
  \bibinfo{journal}{Physics of Life Reviews} \textbf{\bibinfo{volume}{2}},
  \bibinfo{pages}{157} (\bibinfo{year}{2005}).

\bibitem[{\citenamefont{Elgart et~al.}(2011)\citenamefont{Elgart, Jia, Fenley,
  and Kulkarni}}]{elgart2011connecting}
\bibinfo{author}{\bibfnamefont{V.}~\bibnamefont{Elgart}},
  \bibinfo{author}{\bibfnamefont{T.}~\bibnamefont{Jia}},
  \bibinfo{author}{\bibfnamefont{A.}~\bibnamefont{Fenley}}, \bibnamefont{and}
  \bibinfo{author}{\bibfnamefont{R.}~\bibnamefont{Kulkarni}},
  \bibinfo{journal}{Physical Biology} \textbf{\bibinfo{volume}{8}},
  \bibinfo{pages}{046001} (\bibinfo{year}{2011}).

\bibitem[{\citenamefont{Shahrezaei and Swain}(2008)}]{shahrezaei2008analytical}
\bibinfo{author}{\bibfnamefont{V.}~\bibnamefont{Shahrezaei}} \bibnamefont{and}
  \bibinfo{author}{\bibfnamefont{P.}~\bibnamefont{Swain}},
  \bibinfo{journal}{Proceedings of the National Academy of Sciences of the
  United States of America} \textbf{\bibinfo{volume}{105}},
  \bibinfo{pages}{17256} (\bibinfo{year}{2008}).

\bibitem[{\citenamefont{Gillespie}(2001)}]{gillespie2001approximate}
\bibinfo{author}{\bibfnamefont{D.}~\bibnamefont{Gillespie}},
  \bibinfo{journal}{The Journal of Chemical Physics}
  \textbf{\bibinfo{volume}{115}}, \bibinfo{pages}{1716} (\bibinfo{year}{2001}).

\bibitem[{\citenamefont{Jia and Kulkarni}(2010)}]{jia2010posttranscriptional}
\bibinfo{author}{\bibfnamefont{T.}~\bibnamefont{Jia}} \bibnamefont{and}
  \bibinfo{author}{\bibfnamefont{R.}~\bibnamefont{Kulkarni}},
  \bibinfo{journal}{Physical Review Letters} \textbf{\bibinfo{volume}{105}},
  \bibinfo{pages}{018101} (\bibinfo{year}{2010}).

\end{thebibliography}

\end{document}